\documentclass[sigconf]{acmart}

\AtBeginDocument{%
  }

\setcopyright{acmlicensed}
\copyrightyear{2024}
\acmYear{2024}
\acmDOI{XXXXXXX.XXXXXXX}

\usepackage{tabularx}
\usepackage{multirow}
\usepackage{multicol}
\usepackage{threeparttable}
\usepackage{graphicx} 
\usepackage{subcaption} 
\usepackage{soul}

\begin{document}

\title{Understanding the Effect of Algorithm Transparency of Model Explanations in Text-to-SQL Semantic Parsing}

\author{Daking Rai}
\email{drai2@gmu.edu}
\affiliation{%
\institution{George Mason University}
\country{}
}

\author{Rydia R. Weiland}
\email{sweilan@gmu.edu}
\affiliation{%
\institution{George Mason University}
\country{}
}
  
\author{Kayla M. G. Herrera}
\email{kherrera@mitre.org}
\affiliation{%
\institution{MITRE Corporation}
\country{}
}

\author{Tyler H. Shaw}
\email{tshaw4@gmu.edu}
\affiliation{%
\institution{George Mason University}
\country{}
}

\author{Ziyu Yao}
\email{ziyuyao@gmu.edu}
\affiliation{%
\institution{George Mason University}
\country{}
}

\renewcommand{\shortauthors}{Rai et al.}

\begin{abstract}

Explaining the decisions of AI has become vital for fostering appropriate user trust in these systems. This paper investigates explanations for a structured prediction task called ``text-to-SQL Semantic Parsing'', which translates a natural language question into a structured query language (SQL) program. In this task setting, we designed three levels of model explanation, each exposing a different amount of the model's decision-making details (called ``algorithm transparency''), and investigated how different model explanations could potentially yield different impacts on the user experience. Our study with $\sim$100 participants shows that (1) the low-/high-transparency explanations often lead to less/more user reliance on the model decisions, whereas the medium-transparency explanations strike a good balance. We also show that (2) only the medium-transparency participant group was able to engage further in the interaction and exhibit increasing performance over time, and that (3) they showed the least changes in trust before and after the study.

\end{abstract}




\maketitle

\section{Introduction}\label{sec:introduction}

Artificial intelligence (AI) systems are now increasingly used to facilitate human life, such as assisting people in complex daily tasks and boosting their productivity in the workplace. For these systems to be reliably and securely used in critical domains, building appropriate trust between humans and the systems is crucial. Among others, Explainable AI (XAI) has been considered to be a key technique in fostering appropriate human-AI trust by presenting to humans how the AI system reaches a certain decision for a given input (called ``local explanation''), {which has led to the development of several XAI techniques~\cite{ribeiro2016should, scott2017unified, tuan2021local, sundararajan2017axiomatic, camburu2018snli, rai2023explaining}. However, recent human-subject studies suggest that these explanations do not help humans identify AI misclassifications~\cite{carton2020feature, lai2019human}, nor reduce over-reliance ~\cite{zhang2020effect, wang2021explanations, bansal2021does} or foster appropriate human-AI trust~\cite{zhang2020effect}. Nonetheless, most of these studies focus only on simpler classification tasks like sentiment classification or multiple-choice question answering (MCQA), and insights from them may not necessarily generalize to more complex tasks where explanations could provide greater value.}

In this paper, we extend this line of research but seek to examine the effect of XAI on human-AI trust and humans' ability to identify correct/incorrect predictions in a more complex task, called ``text-to-SQL semantic parsing''~\cite{zhong2017seq2sql, yu-etal-2018-spider, xie2022unifiedskg, li2024can}. {Text-to-SQL semantic parsing involves translating a question written in natural language (e.g., \textit{``How many schools or teams had jalen rose?''}) to a Structured Query Language (SQL) query (e.g., \textit{SELECT COUNT(School/Club Team) FROM Table1 WHERE Player = ``jalen rose''}), which can then be executed against the provided database to retrieve relevant database records or calculate the queried results in the natural language questions.} Unlike the classification tasks that were commonly used for case studies in prior XAI work, semantic parsing represents a much more complicated ``structured prediction'' problem, where a model needs to make a sequence of inter-correlated predictions, each corresponding to one token constrained by the grammar of the target formalism, to form the final logical form (e.g., a SQL query). The complexity of this task thus makes the application of XAI non-trivial, giving rise to two critical questions: \emph{(1) How to explain a semantic parsing result?} Prior work explained an AI model's classification result by offering one explanation for each label, elaborating on why AI thinks each one of them could be the correct answer, and participants were required to select the label with the most convincing rationale or explanation~\cite{wang2021explanations, bansal2021does, zhang2020effect}. However, these prior approaches cannot be directly applied to explaining semantic parsing consisting of sequential decision-making and numerous possible outputs. 
\emph{(2) How do various model explanations affect the user experience, particularly their utilization of AI and their trust in the model?}
Given the complexity of the semantic parsing task, the model explanations for the semantic parser can be designed with varying levels of detail about its decision-making process (called ``algorithm transparency levels''), and it is unclear how these explanations influence the user experience. In our work, we particularly focus on two aspects of impact, i.e., whether these explanations allow users to recognize correct vs. incorrect AI predictions and utilize AI for their best benefit, and whether they lead to enhanced or dampened trust after users interact with the model for some tasks. We note that this is not a trivial question. For example, when more information about the model is exposed (i.e., providing {high-transparency} explanations), users may benefit from the rich information and thus make a better judgment on the correctness of the model predictions. However, the rich explanations may also be too complicated for users without sufficient backgrounds to digest; as a result, users may suffer from confusion and thus reduce their trust in the model.

To thoroughly understand the effect of model explanation on user experience in the task of text-to-SQL semantic parsing, we conducted a human-subject study. We experimented with three distinct explanation approaches at low, medium, and high levels of algorithm transparency, more details in Section~\ref{sec:transparency-level-design}. We recruited 97 participants who had no background in computer science or SQL programming, to emulate the situation when the semantic parser is applied to assist non-technical users (e.g., administrative staff in schools and companies) in accessing database information. Each participant was instructed to complete 30 text-to-SQL tasks with assistance from a language model-based semantic parser~\cite{xie2022unifiedskg} and was then prompted to decide if the semantic parser's prediction was correct based on the presented model explanation. We also measured their trust in the semantic parser before and after the tasks using the Propensity to trust~\cite{merritt2013trust} and Jian scale trust measurement~\cite{jian2000foundations}, which are two commonly adopted metrics in Psychology field studies. 

Our analysis revealed multiple interesting findings. 
\emph{First,} surprisingly, no obvious impact was observed from the algorithm transparency level of a model explanation on the participant’s overall success rate in distinguishing between incorrect and correct model predictions, although they do lead to different user behaviors (e.g., the high-transparency group seemed to be persuaded by the informative explanations and tended to accept more predictions, while the low-transparency group showed the opposite).
Interestingly, we observed a declining trend in the participant performance when they interacted with the AI model for a longer time. The exception happened to only participants receiving the medium-transparency level of model explanations; as time went by, these participants gradually adapted themselves to this explanation method and obtained increasing performance in distinguishing between incorrect and correct model predictions.
\emph{Second,} as we expect, participants spent significantly more time reading model explanations at the high transparency level than others. Observing no strong effect from high-transparency model explanations on participant performance (as discussed previously) may indicate that the participants, due to a lack of SQL programming background, were not able to digest the full details of the model decision-making. Interestingly, we observed a degrading participant performance 
when they spent more time reading the low-transparency model explanations. This was likely caused by that the participant, when presented with only minimal information, invented their own strategy of judging the model decisions, and this self-developed strategy was not only time-consuming but also inaccurate.
\emph{Finally,} {while all participants showed a decline of trust in the Jian scale trust measure after the study, participants receiving medium-transparency explanations showed the least change in the trust level. This observation may imply a key consideration when future researchers design the model explanation method. }That is, when the AI application is complex, including either too little (i.e., low-transparency level) or too much (i.e., high-transparency level) information about an AI model will only hurt the human-AI trust. Instead, the proper amount of algorithm information should be decided based on the prospective users' backgrounds.

\section{Background and Related Work} \label{sec:background}

\subsection{Explainable AI and Trust in Human-Machine Interaction}
Formally, ``trust'' is understood as ``an attitude of confident expectation in an online situation of risk that one’s vulnerabilities will not be exploited''~\cite{corritore2003line}. Prior work~\cite{de2020towards} has identified two cases detrimental to the interaction between humans and machines: (1) Over-trust, i.e., when humans trust ``too much'' in the machine, which could lead to ``hands-off'' monitoring behavior rendering the human unable to respond to an error or malfunction; (2) Under-trust, i.e., when humans trust ``too little'' in the machine, even when the machine has outstanding capabilities in tasks, which could cause an unnecessarily unbalanced workload and inefficient collaboration. Establishing proper human trust in the AI model is thus crucial for secure human-machine interactions~\cite{de2018automation}. Recent research revealed that controlling ``algorithm transparency'', i.e., the amount of algorithm detail to be exposed, could be a promising solution to calibrate human trust in machines~\cite{kizilcec2016much}. However, this idea has not been studied in a task setting as complicated as semantic parsing.

In connection with algorithm transparency, how to effectively explain an algorithm or a model has been a long-standing problem, giving rise to the research topic of Explainable AI (XAI)~\cite{molnar2020interpretable}. One common scenario of explanation is to locally explain why a model gives a certain output given the input (i.e., ``local explanation''). While this way of explanation does not offer much global information about the model's intrinsic properties, it enjoys the benefit of being targeted to the individual model decision. Feature attribution is one type of method for local explanation. It explains a model by showing which input features the model output should be attributed to. Some well-known feature attribution-based explanation methods include LIME~\cite{ribeiro2016should}, Shapley value~\cite{shapley1953value}, Kernel SHAP~\cite{scott2017unified}, and Integrated Gradients~\cite{sundararajan2017axiomatic}. Our study used LERG~\cite{tuan2021local}, a recent feature attribution method designed for conditioned generation tasks, where the model outputs a sequence of tokens conditioned on the input. LERG adapted Shapley value and LIME into LERG-S and LERG-L; we used LERG-S in our medium- and high-transparency explanations.

Many prior studies have similarly examined the effect of model explanations on user experience or perception~\cite{doshi2017towards, narayanan2018humans, hase2020evaluating, bansal2021does, shen2022shortest, yao2023human}. \citet{doshi2017towards} defined a taxonomy of interpretability evaluation, consisting of application-grounded evaluations involving real humans and real tasks, human-grounded metrics involving real humans but simplified tasks, and functionally-grounded evaluations involving no humans and only proxy tasks. Our study falls under the first category. Similar to us, \citet{narayanan2018humans} investigated how increasing the complexity of model explanation could have an impact on the time that humans would need to judge its rationale. Their work considered three variables, namely the explanation size, the new cognitive chunks, and the use of repeated terms. Through human subject studies in a synthetic task across two domains, the authors found that increasing the explanation complexity (e.g., increasing the explanation size or adding new cognitive chunks) generally led to a longer response time for the participants; however, these variations did not significantly affect the human accuracy in validating the correctness of the model decision, similar to what we found in our study. {Additionally, several prior works have investigated the utility of explanations in AI-assisted decision making, aiming to optimize human + AI team performance~\cite{bansal2021does, wang2021explanations}. These studies indicate that although human + AI teams typically outperform individuals working alone, their performance often falls short of that of the AI alone. This inferior performance is frequently attributed to over-reliance on AI, where humans, instead of combining their insights with an understanding of the AI's decision-making process, tend to follow the AI's suggestions even when they are incorrect. To address this issue, \citet{buccinca2021trust} examined whether compelling participants to engage more thoughtfully with AI-generated explanations could reduce over-reliance in collaborative decision-making. They discovered that cognitive forcing mechanisms, techniques designed to make participants think more critically about the AI's suggestions, significantly reduced over-reliance compared to simpler explainable AI methods. However, they also found that participants preferred and trusted systems perceived as less mentally demanding, even if their performance with such systems was lower.}
Despite all the prior effort, we note that these earlier works have mainly focused on the simplified classification tasks, while the effect of variations of explanations has not been examined in a task as complicated as semantic parsing.

\subsection{Explaining Text-to-SQL Semantic Parsing}
Our project is based on a structured prediction task called ``text-to-SQL semantic parsing''. A text-to-SQL semantic parsing system aims to translate a natural language text 
into a SQL query,
enabling non-technical users without SQL programming skills to interact with databases and retrieve information using natural language. As a result, the past few years have witnessed continuing excitement of building text-to-SQL semantic parsers~\cite{zhang2019editing, lin2020bridging, Scholak2021:PICARD, xie2022unifiedskg, rai2023improving, li2023resdsql, pourreza2024din}. In our work, we used a semantic parser built from the open-source T5 language model~\cite{raffel2020exploring}, following the same approach of \citet{Scholak2021:PICARD, xie2022unifiedskg, rai2023improving}.

While machine performance on this task has been boosted dramatically in the past years, state-of-the-art systems still fall short in real applications, due to practical challenges such as language ambiguity or complexity and domain shift. This inspires a recent line of research called ``interactive semantic parsing'', where the semantic parsing system proactively explains its decisions to the human and seeks human feedback to correct potential mistakes~\cite{gur2018dialsql, yao2019model, elgohary2020speak, zeng2020photon, li-etal-2020-mean, 10.1145/3397481.3450667, tian2023interactive}. Our work was inspired by the rise of this line of research but was focused exclusively on the impact of various explanations on the end users of the semantic parser. We aim to examine this impact systematically with a human subject study, which was barely performed by prior works.

Our work involves three distinct explanation methods, i.e., model confidence, feature attribution, and visualized step-by-step explanations. These methods are considered representative ways of model explanations and have been explored by prior works. \citet{dong2018confidence} were among the earliest in modeling a semantic parser's confidence in its prediction. In their work, the authors categorized a semantic parser's uncertainty into three types, i.e., model uncertainty, data uncertainty, and input uncertainty, and developed approaches to measure each type respectively. \citet{yao2019model} followed a similar approach as \citet{dong2018confidence} but found that the confidence score might not be a good indicator of whether a semantic parser made a correct or incorrect prediction. Inspired by the need for a more reliable indicator, \citet{stengel2023calibrated} studied approaches for calibrating a semantic parser's confidence score.
In the spectrum of text-to-SQL semantic parsing, feature attribution is largely understudied. The only available is that of \citet{rai2023explaining}, which systematically compared different feature attribution approaches (e.g., LIME, Shapley, and LERG). However, the comparison did not involve any human subjects; instead, it was mainly based on automatic evaluation metrics, yet whether these metrics represent user perception of explanations in real life was uncertain. 
Finally, the step-by-step explanation in our high-transparency explanation design was inspired by prior works of \citet{elgohary2020speak, 10.1145/3397481.3450667, tian2023interactive}, which demonstrated the effectiveness of this approach in text-to-SQL semantic parsing. In particular, \citet{10.1145/3397481.3450667} proposed to include dynamic views of database changes when a SQL query is executed step by step. We borrowed this idea in our high-transparency explanation design as well. However, we note that none of the prior works have systematically compared all the three types of explanations, especially when they are organized to represent different algorithm transparency levels, in one human subject study. 
\section{Study Goals}

Our study aims to understand whether and to what extent a model explanation at a different level of algorithm transparency will impact the user experience when they interact with the AI model. We define different algorithm transparency levels based on the amount of model detail that the explanation exposes to the user, such as showing the entire or only partial decision-making process of the model. Our goals for this study are to answer the following two research questions (RQs):
\begin{itemize}
    \item \textbf{RQ1: How do the model explanations at different transparency levels affect humans' ability to accurately identify correct and incorrect AI predictions?} We follow the same spirit of prior works~\cite{zhang2020effect} in considering how the model explanations can assist humans in better utilizing the AI model and maximizing their benefit in tasks. Specifically, we focus on the scenario where users are presented with the model prediction, but will make a judgment on whether the prediction is correct or not, based on the model explanation and other task information. An effective explanation should allow users to accurately distinguish between correct and incorrect AI predictions, such that they can avoid the risks of adopting incorrect predictions from the AI model. This is a non-trivial RQ in our setting because of the pros and cons of each transparency level of explanation. For example, while a high-transparency explanation could allow users to make more informed judgments owing to the larger amount of model detail it provides, it may also confuse the users if the users do not have the proper knowledge background to digest the information. In contrast, while a low-transparency explanation only provides limited insights about a model's decision, the low cognitive load it requires may engage the users better. In our study, we hope to discuss these trade-offs carefully from the empirical observations of participant performance.
    \item \textbf{RQ2: How do the model explanations at different transparency levels affect humans' trust in the AI model?} Similarly, our study also aims to understand the impact of explanations at different transparency levels on human-AI trust. As RQ1, this is not a trivial question either. The very first challenge lies in the measurement of human trust. While ``trust'' has been discussed in many prior works which similarly investigated in the effect of model explanations~\cite{zhang2020effect, ferrario2022explainability}, there is no clear strategy for effectively measuring human trust. Second, whether an explanation at a certain transparency level will lead to more or less human trust in AI, is a complicated question. For example, a high-transparency explanation could increase human trust when it allows the user to know more about the model and hence enhances their confidence in using the model, but it could also lead to less human trust when the explanation is not plausible from the human perspective, although it is faithful to the model itself.
\end{itemize}

While Explainable AI (XAI) and its effect on user experience have been explored by multiple prior works~\cite{wang2021explanations, bansal2021does, zhang2020effect}, most, if not all, of these works, focused on simplified classification tasks, such as sentiment classification and multiple-choice question answering (MCQA), when there are only a limited set of labels for the AI model to pick. However, the space of machine learning and AI includes way more complicated tasks than classification, and these tasks do not come with a finite set of possible answers. This discrepancy significantly increases the difficulty in effectively explaining these complicated AI models and renders findings from prior literature not directly applicable. {For instance, in sentiment classification or MCQA tasks, explanations for all possible labels can be presented to participants, who then compare them and choose the most convincing explanation. In contrast, for tasks with numerous possible outputs, it is impractical to provide one explanation for each potential answer. Typically, explanations are generated only for the AI's predicted answers, which may not always be correct. Consequently, participants must decide whether to trust the AI based solely on the explanation of the generated output, which is more challenging. Additionally, the complexity of these tasks often requires domain-specific knowledge, further complicating the participants' ability to evaluate the explanations. These factors can result in varying participant behaviors and attitudes toward the explanations provided.}

In our study, we pick ``text-to-SQL semantic parsing'' as one of such complicated AI tasks, and use it to examine the two RQs we defined above. This is a task of automatically converting a natural language question to a SQL query, such that by executing the SQL query against the given database, one can obtain the database querying results expressed in their natural language question~\cite{xie2022unifiedskg}. By its nature, semantic parsing is a structured prediction task, as its goal is to generate a sequence of code tokens forming the structure of a grammatically correct SQL query. Unlike classification, semantic parsing has an infinite label space; that is, in principle, a semantic parser can generate an arbitrary number of SQL programs. On the other hand, users of semantic parsers are often non-technical people who cannot write a SQL program themselves, because otherwise they can directly compose the SQL query without needing help from the semantic parsers. This particular user group also makes explaining a semantic parser's SQL prediction difficult. Below, we further highlight the complexity of this task and the challenges in interpreting its study results, which will serve as a foundation for our result analysis.

\paragraph{The substantially higher complexity of semantic parsing than simple classification.} While in classification tasks a user mainly needs to read the text input (e.g., the sentence to be predicted with a sentiment label), in semantic parsing tasks, multiple components will be involved. Specifically, as we will describe in Section~\ref{sec:interface-design}, 
a user interacting with a semantic parser needs to first understand the database based on its schema and the provided sample records, and then interpret their question by grounding it onto the database.
Although users are most concerned with the final received records (e.g., for \emph{``what/which''} questions) or the calculated results (e.g., for \emph{``how many''} questions), the semantic parser functions to generate the SQL query, not to directly retrieve the records. Therefore, explaining the model's decision-making process requires us to clarify both the generated SQL and its interaction with the database to retrieve the records. Designing such an informative explanation
without significantly increasing the cognitive load on the participants can be challenging. Finally, we note that, despite its complexity, semantic parsing offers an effective and a unique task setting to analyze the impact of explanations at varying levels of algorithm transparency on participants' performance and trust levels.

\paragraph{Complexity caused by the users' non-technical backgrounds.} 
As discussed above, explaining the AI's decision-making process often involves detailing the predicted SQL query. Users without a technical background may find this information overwhelming, particularly when dealing with complex SQL queries that contain numerous clauses. This gap of knowledge could affect their understanding of the task, the model predictions, and the model explanations, which eventually impacts their performance in recognizing correct and incorrect AI predictions and their trust levels. In our study, we also aim to analyze users' changes of behaviors, such as when these non-technical users keep interacting with such a complicated system, whether they will actively learn from the model explanations and gradually adapt themselves to tasks at this complexity level.

\paragraph{The intricate relation between RQ1 and RQ2.} 
The user performance in recognizing correct and incorrect predictions (RQ1) and their trust in the AI model (RQ2) are not the same. 
Rather, the user performance is often \emph{an effect of multiple factors, including their trust in the AI model.} Specifically, whether a user can correctly distinguish between correct and incorrect predictions could depend on their trust in the model, their knowledge background (e.g., how much they can understand about the task and the model explanations), their personality (e.g., whether they tend to be patient in reading very long descriptions as in the high-transparent explanations), and the quality of the explanations (e.g., whether the explanations are precise and indicative enough for anyone to make a reasonable judgment). Assume that the AI model has an accuracy of $P$ in the semantic parsing task. We consider the following three cases when connecting the user performance in RQ1 with their trust measured in RQ2:
\begin{itemize}
    \item In the case of \emph{extreme} under-trust, users may simply reject \emph{any} predictions from the model and consider all of them as incorrect. In this case, their accuracy in successfully recognizing correct and incorrect predictions will be $1 - P$.
    \item In the case of \emph{extreme} over-trust, users simply accept \emph{any} predictions and consider them as correct. Their accuracy in this case will be the same as model accuracy, i.e., $P$.
    \item In the case of normal trust, users not necessarily can obtain high or low accuracy. As we elaborated above, this accuracy depends on multiple factors. In an extreme case when all the factors are perfectly set (e.g., model explanations are perfectly designed, users are reliable in reading long explanations and completing the task carefully, etc.), the user accuracy will be 100\%, which shows the best situation. However, in reality, there are always factors that are not perfectly set, which results in uncertainty in user accuracy, and the final accuracy could be larger or smaller than the accuracies in the extreme under-trust and over-trust cases.
\end{itemize}
The intricate relation between RQ1 and RQ2 makes the interpretation of our study results challenging yet interesting.
\section{Interface Design for Human Subject Study} \label{sec:interface-design}

\begin{figure*}[t!]
    \centering
    \includegraphics[width=0.7\linewidth]{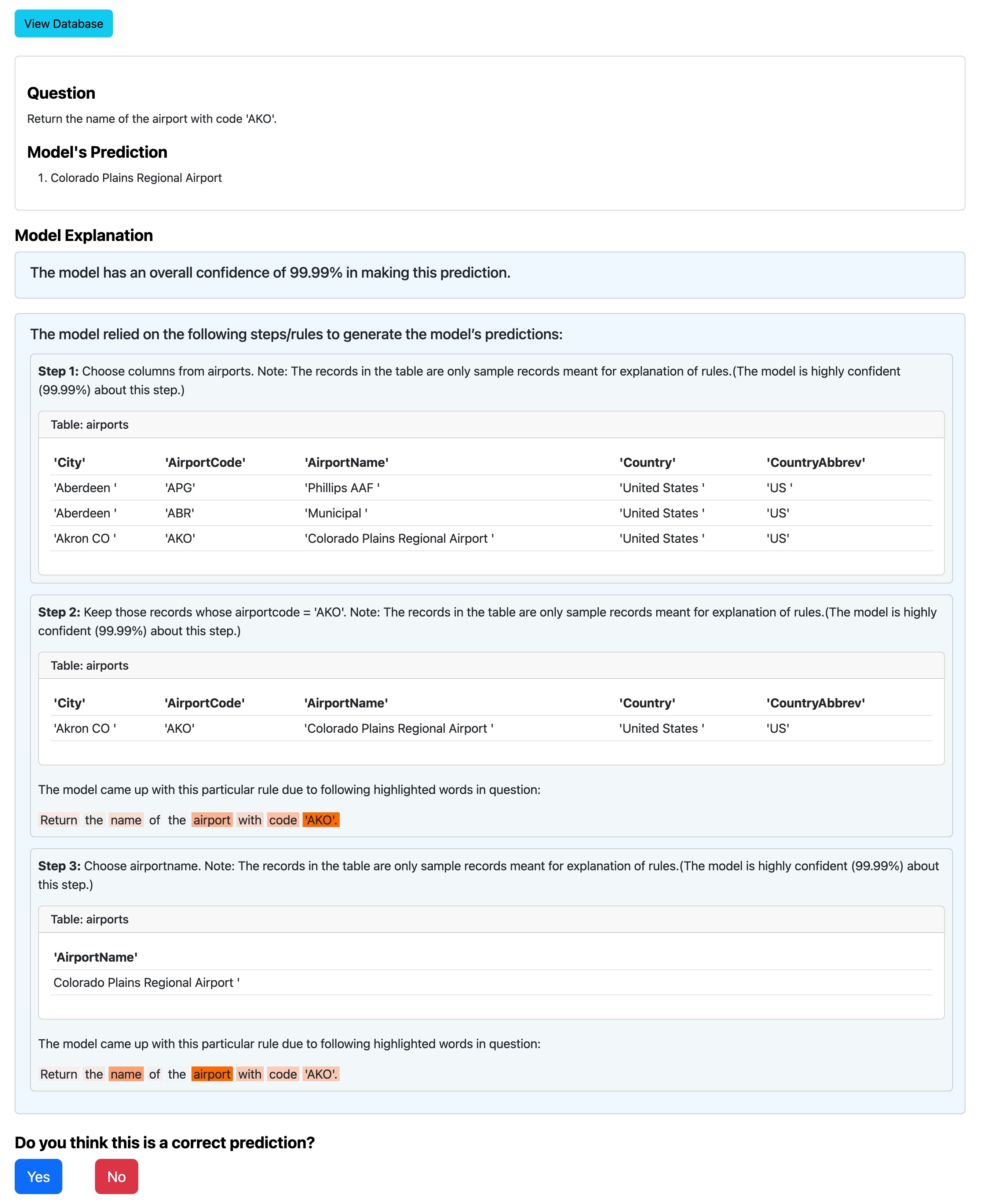}
    \caption{Interface designed to facilitate our human subject study with high-transparency explanation.}
    \label{fig:interface}
\end{figure*}

\begin{figure*}[th!]
    \centering
    \includegraphics[width=0.7\linewidth]{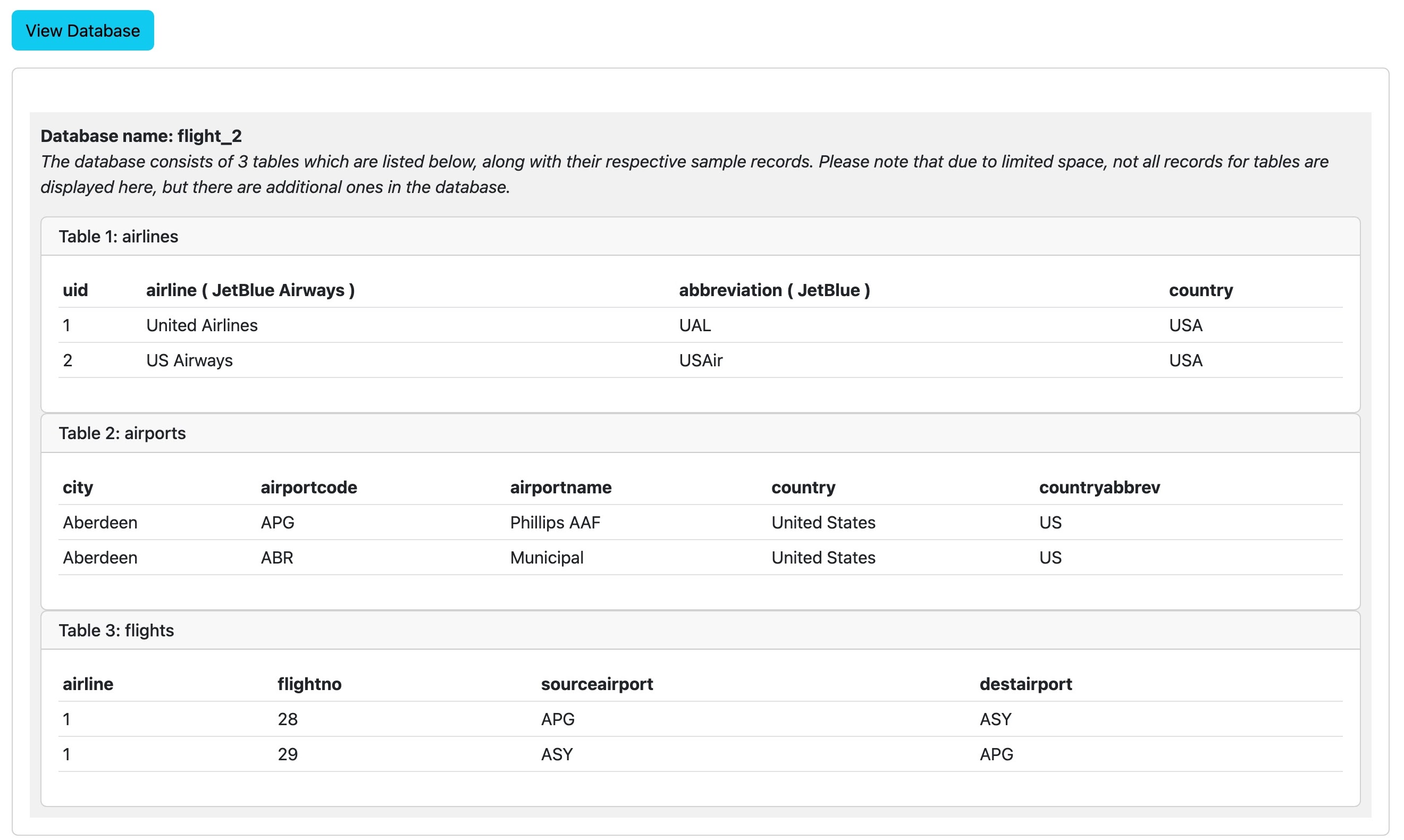}
    \caption{The ``database view'' in our interface presents the participants with the database schema and sample table records.}
    \label{fig:database-view}
\end{figure*}

\begin{figure*}[th!]
    \centering
    \includegraphics[width=0.7\linewidth]{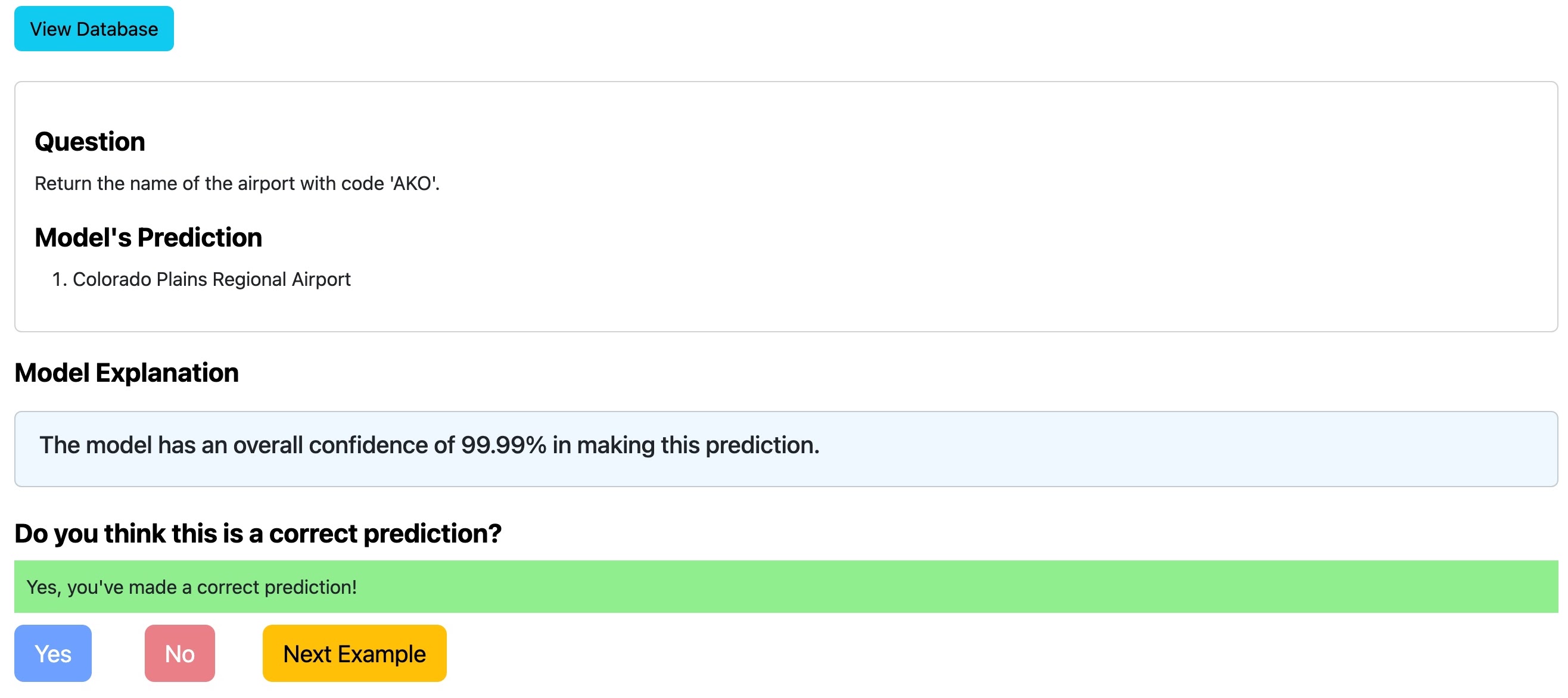}
    \caption{A message will pop up after the participant chooses an answer to the question ``Do you think this is a correct prediction?''.}
    \label{fig:pop-up-msg}
\end{figure*}

We designed a user interface (UI) to support our human subject study, as shown in Figure~\ref{fig:interface}. Specifically, each participant will first be presented with a natural language \textbf{Question} describing the database query task they need to complete. The specific database context can be viewed by clicking the \textbf{View Database} button, which includes the name of the database, the schema and the name of each table within this database, as well as a few sample records from each table (Figure~\ref{fig:database-view}). Note that we intentionally provide only the record samples to simulate the practical situation when humans cannot fully explore a large-scale database due to limitations on time and effort; instead, they can typically browse the first few lines of records to form a rudimentary understanding of the database information. After the question and the database view, we present the \textbf{Models' Prediction} results, which are obtained by executing the model-generated SQL query against the complete database. Note that the participants are assumed to have no SQL programming skills, so they will not understand the model-predicted SQL. Here, we only present the participants with the retrieved database records or the calculated results from executing the model-generated code.

Following these items, we present a \textbf{Model Explanation}, which will be implemented with various approaches to explain the model prediction to the participant, as we will introduce in Section~\ref{sec:transparency-level-design}. To measure the effect of each model explanation, we then prompt the participant to judge, based on the explanation, if the model's prediction is correct or not. This prompt, ``Do you think this is a correct prediction'', is included at the bottom of the UI.
Ideally, an effective model explanation should allow participants to distinguish between correct and incorrect model predictions.
After the participant clicks one of the choices, a message will pop up showing them the true judgment (Figure~\ref{fig:pop-up-msg}). The purpose of this pop-up message is to {provide immediate feedback on whether their judgment was accurate, allowing them to utilize the explanation more effectively for subsequent questions}.As a result, we also expect the participants to learn from past interactions and gradually adapt themselves to the task with increasingly better performance.

\section{Three Levels of Explanations for Text-to-SQL Semantic Parsing} \label{sec:transparency-level-design}

In our study, we experimented with three explanation approaches, each at a distinct algorithm transparency level. 

\begin{figure*}[th!]
    \centering
    \includegraphics[width=0.7\linewidth]{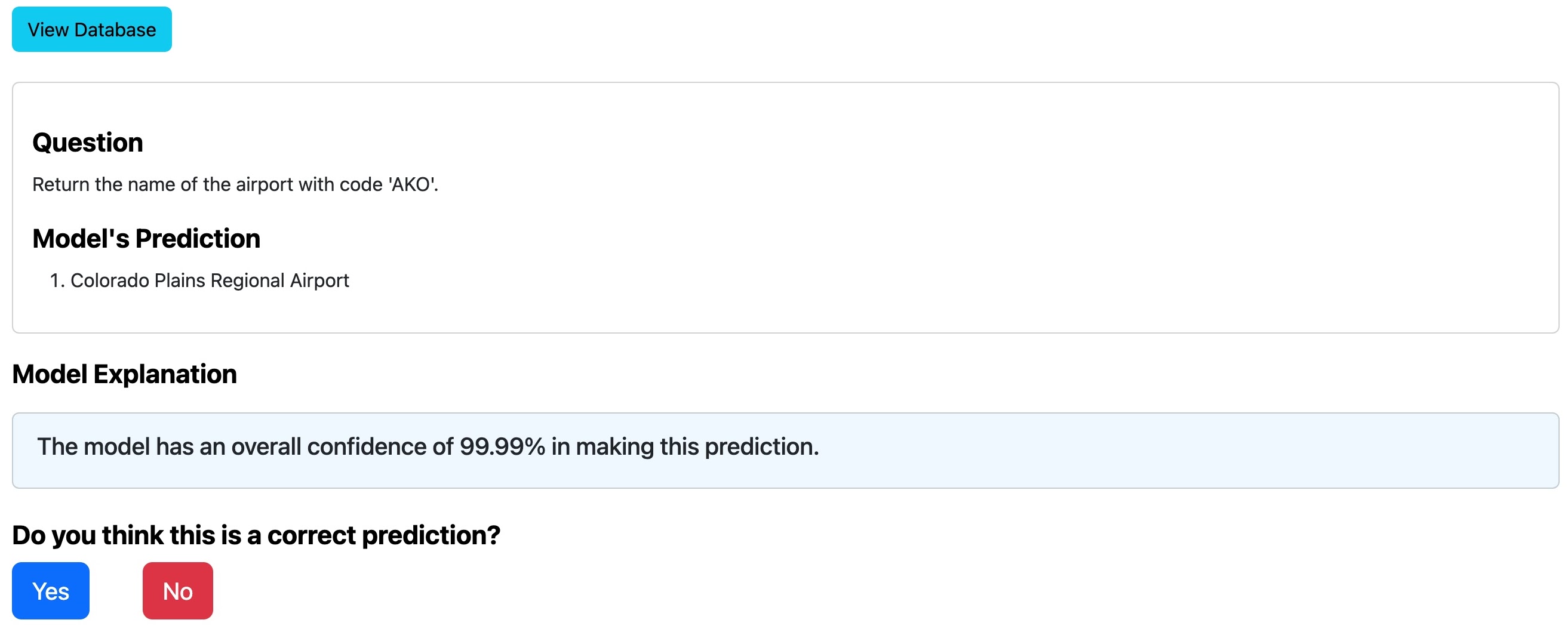}
    \caption{An example explanation at a low transparency level.}
    \label{fig:low-level}
\end{figure*}

\textbf{Low-transparency Explanation} (Figure~\ref{fig:low-level}) presents to participants only a confidence score of the model's decision-making (i.e., the SQL query prediction), which is calculated by taking the arithmetic average of the softmax probabilities assigned to each token in the SQL query by the model during prediction. Specifically, the confidence score $C$ for the generated SQL query with $N$ tokens is given by $C = \frac{\sum_{n=1}^{N} c_n}{N}$, where $c_n$ is the confidence score for the $n$-th token in the predicted SQL query, which is calculated as the model's conditioned probability of generating this token, i.e., $c_n = P(c_n|c_1,\cdots,c_n)$.

\begin{figure*}[th!]
    \centering
    \includegraphics[width=0.7\linewidth]{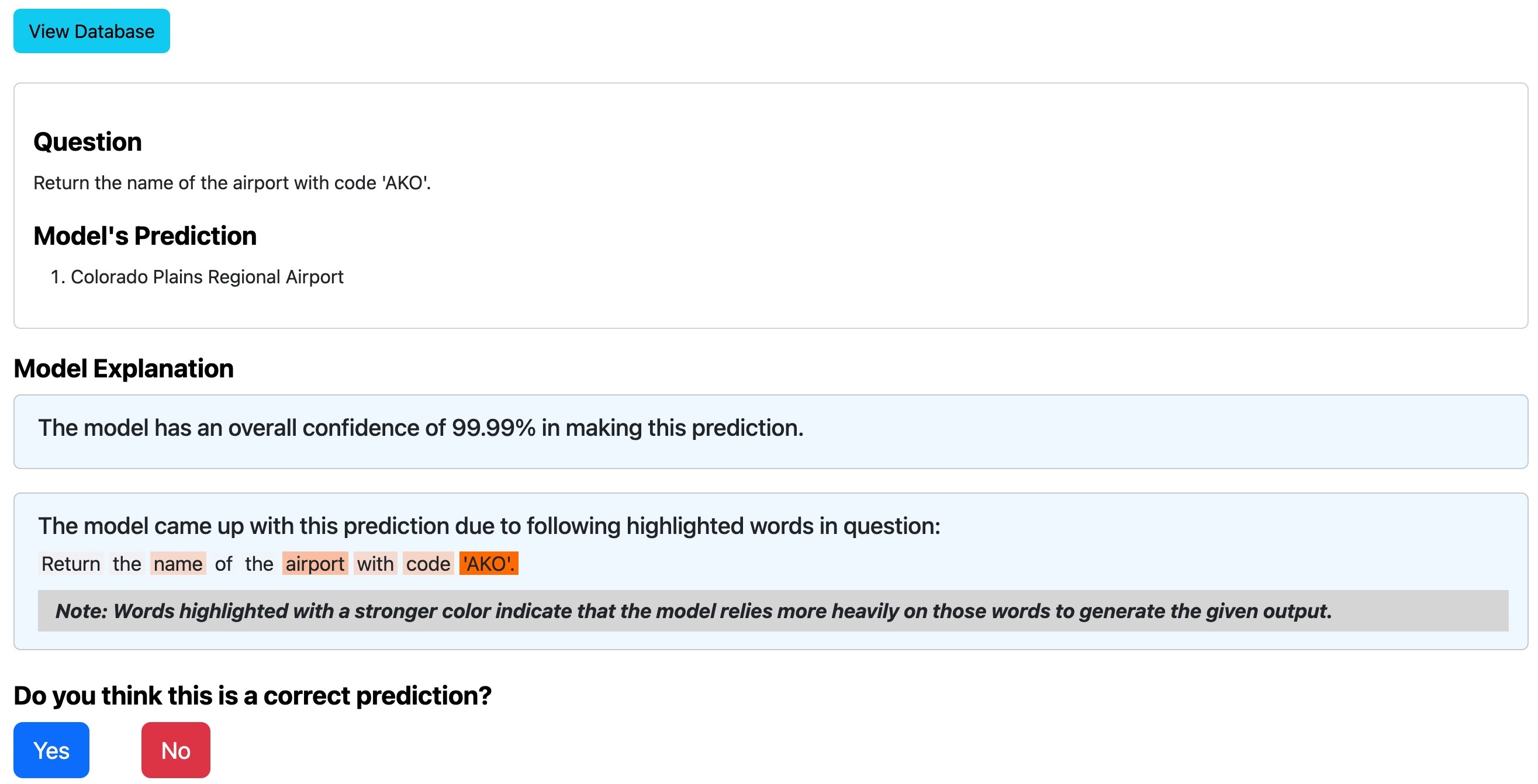}
    \caption{An example explanation at a medium transparency level.}
    \label{fig:mid-level}
\end{figure*}

\textbf{Medium-transparency Explanation} presents to participants both the confidence score of the generated SQL query and a feature attribution explanation, highlighting the input features considered most important by the model. For instance, the example in Figure~\ref{fig:mid-level} shows that the model decides to return the retrieved database records mostly because of the words ``name'', ``airport'', ``code'', and ``AKO'' present in the input question, ``Return the name of the airport with code 'AKO'.''. Note that other words also have some impact on the model’s decision, but they are very light and negligible. The contribution of features to the model prediction was extracted using the LERG-S algorithm~\cite{tuan2021local}, a local explanation method that adapts Shapley value~\cite{shapley1953value} to explain models in generation tasks.

\textbf{High-transparency Explanation} provides participants with the overall confidence score for the predicted SQL query, along with a detailed explanation of the SQL itself, as shown in Figure~\ref{fig:interface}. Specifically, the SQL was first translated to NatSQL~\cite{gan2021natural}, an intermediate representation that simplifies the SQL to make it easier to identify text descriptions for each clause. Each clause was then explained in plain English as a step or rule, to help participants understand how the query interacts with the database to generate the final results. For example, Figure~\ref{fig:interface} shows a SQL query composed of three rules for generating the answer to the question: first, it retrieves all records from the \emph{airports} table; then, it filters the results to include only those whose \emph{airportcode} is ``AKO''; and finally, the SQL query returns values from the \emph{airportname} column. Additionally, the confidence score and feature attribution explanation for each rule were also displayed, helping participants understand how confident the AI was in predicting each rule and which input features it considered significant for generating the rule. {Note that the feature attribution explanation for the first step/rule is absent because NatSQL does not have a separate FROM clause; we however included the description and database view of this step to complete the explanation.} This step-by-step breakdown provides a comprehensive view of all the SQL query rules, explains why the model generated each one, and shows how they were used to obtain the final answer.

To summarize, our study involves three types of model explanations designed for the complicated text-to-SQL semantic parsing task, each representing one algorithm transparency level. We also note that in our design, information included in the lower transparency level of explanation is a subset of information in the higher level one. For example, the model confidence score is included in all three levels, and when the feature attribution results are included in the medium-transparency explanations, they are also displayed in the high-transparency explanations. As such, results from our study allow us to easily understand the effect when the amount of information increases from level to level.

\section{Trust Measurement} \label{sec:trust-measurement}

Trust is a multi-dimensional construct~\cite{mayer1995integrative, lee2004trust, hoff2015trust, kohn2021measurement}. For example, \citet{hoff2015trust} posit that trust consists of dispositional trust (an individual's overall tendency to trust automation independent of context), situational trust (how trust changes in response to context), and learned trust (an individual's evaluation of a system drawn from past experience). Due to its multi-dimensional nature, authors have suggested that trust should be measured in a variety of ways~\cite{kohn2021measurement}. For this reason, we measured dispositional trust with the ``propensity to trust scale''~\cite{merritt2013trust} and learned trust with the ``checklist for trust'', or the Jian trust scale~\cite{jian2000foundations}. Below, we explain the two trust measures further.

\paragraph{Propensity to Trust.}
The Propensity to Trust questionnaire developed by \citet{merritt2013trust} was used to measure dispositional trust in automated systems. This 6-item scale was developed to assess the broad, trait-like tendency to trust machines. It involves Likert ratings on a scale from 1 (strongly disagree) to 5 (strongly agree). Sample questions include \emph{``I usually trust machines until there is a reason not to''} and \emph{``For the most part I distrust machines''}. We note that in this questionnaire, only the second item ask about negative opinions while the remaining are all positive. Therefore, participant responses to these two subsets of items should be interpreted differently. For example, when a participant gives a score of 1 (strongly disagree) to a negative item \emph{``For the most part I distrust machines''}, they indeed mean a very positive attitude to machines, but a score of 1 for a positive item \emph{``I usually trust machines until there is a reason not to''} implies the opposite.

\paragraph{Checklist for Trust between People and Automation (Jian Scale).}
The Checklist for Trust developed by \citet{jian2000foundations} was used to measure trust between people and automated systems. It involves Likert ratings ranging from scores 1-7 to 12 items. Similar to the propensity to trust measurement, the checklist includes both positive and negative items. Specifically, The first 5 items check humans' negative opinions toward the automated system; sample questions include \emph{``The system is deceptive''} and \emph{``I am wary of the system''}. The remaining 7 items check humans' positive opinions toward the system, including questions such as \emph{``I'm confident in the system''} and \emph{``The system provides security''}.

\section{Method} \label{sec:human-study}

\subsection{Task Setup}\label{subsec:task}

\begin{figure}[t!]
    \centering
    \includegraphics[width=0.9\linewidth]{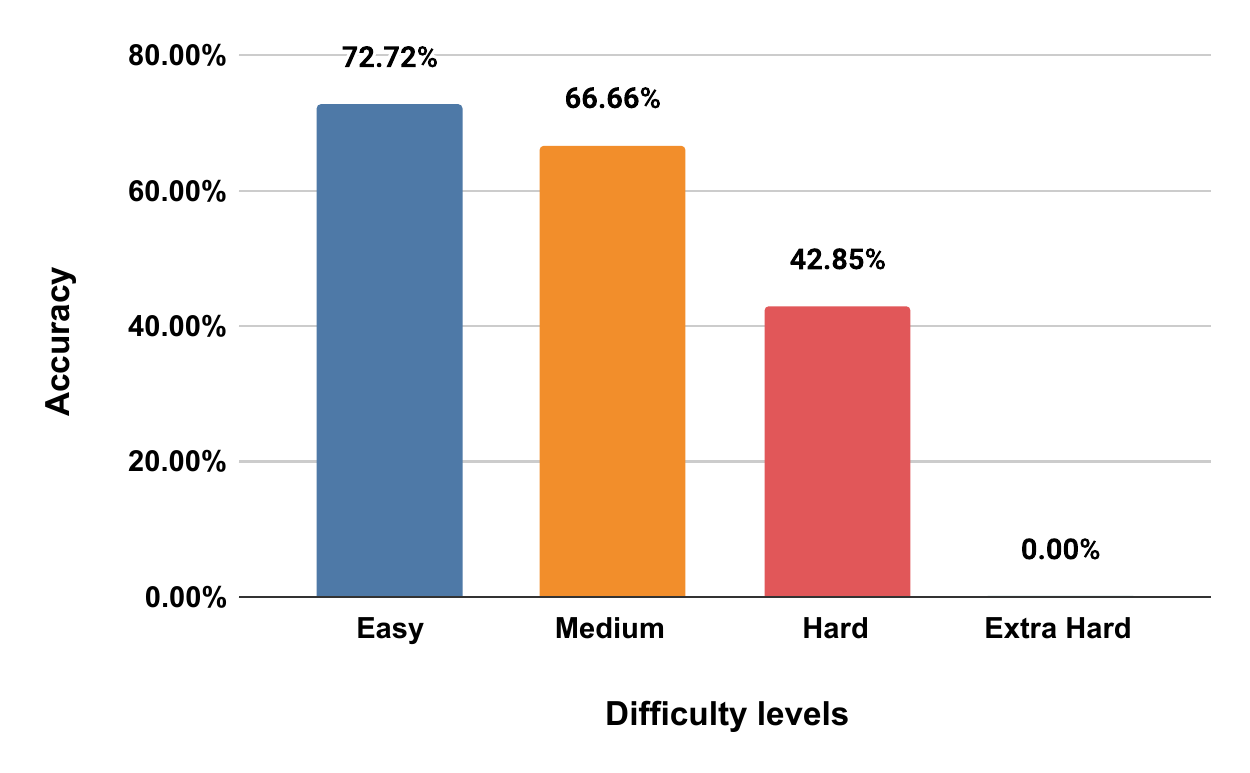}
    \caption{The accuracy of the fine-tuned T5-base semantic parser across sampled 30 examples at different difficulty levels.}
    \label{fig:semantic-parser-accuracy}
\end{figure}

We fine-tuned an encoder-decoder language, T5-base~\cite{raffel2020exploring}, to be our semantic parser. An advantage of this model architecture lies in its mediocre performance, which provided us with a balanced set of samples where the model made both correct and incorrect predictions. We followed prior works~\cite{Scholak2021:PICARD, xie2022unifiedskg, rai2023improving} in formulating the input to the T5 model consisting of the natural language question and the database information, and the output from the T5 model being the SQL query.
Both the model fine-tuning and our study used the Spider dataset, a large-scale, complex, and cross-domain dataset that has been considered a standard benchmark for text-to-SQL semantic parsing~\cite{yu-etal-2018-spider}. The fine-tuned T5-base semantic parser had an execution accuracy of 57.9\% and an exact match accuracy of 57.2\% on the Spider evaluation set (the development-set split). For the user study, we randomly sampled 30 examples from the Spider evaluation set, where 17 examples were correctly predicted by our semantic parser, reflecting an actual accuracy of 17/30 or 56.67\%. The examples covered all difficulty levels (Easy: 11, Medium: 9, Hard: 7, Extra Hard: 3) following Spider's standard, which defines the difficulty level of an example based on the number of keywords, components, and clauses in its ground-truth SQL query.
Figure~\ref{fig:semantic-parser-accuracy} shows the distribution of the model's accuracy
across these difficulty levels.

\subsection{Participant Recruitment}
To recruit participants without technical backgrounds to meet our study goals, we collaborated with the Psychology department in the institution, which is also the organization of co-authors in this paper. Specifically, we used undergraduate students as our potential participants. These students did not have any programming background. They were compensated for 1 research credit that could be applied to meeting the requirements of their degree program. In total, 115 students responded to our recruitment, who were then invited to complete tasks following our study procedure (to be introduced next). Among them, 18 left the study incomplete, possibly due to them being unable to understand the study or having other personal difficulties. This eventually gives us 97 valid data points for our subsequent analysis. These 97 participants have the following demographics: {aging between 18 and 41 years old; 41 males, and 56 females; and from different ethnic backgrounds (36 Whites, 23 Asians, 10 Black, 8 Hispanic, 4 Middle Eastern, and 16 others).}

\subsection{Study Procedure}\label{subsec:procedure}
Each participant was initially assigned to one of the three transparency-level groups at random. This assignment was unknown to the participants.
The study was conducted virtually on a Web platform with a confidential login developed by the institution's Psychology department to support human subject research. When participants worked on the study, the platform was displayed in a full-screen mode protecting the participants from being distracted by irrelevant outside events. The participants were first presented with an informed consent form, clarifying the research goals and procedures of the study, as well as any potential risks and benefits. Only participants who signed the consent form moved on to the next step. They were then prompted to provide their demographic information and complete the two trust measurements, i.e., the propensity to trust and the Jian Scale measurements. These two metrics assessed the participant's initial trust level.
Before starting the formal study, participants went through a training session identical to the formal study itself, which included 10 examples randomly sampled from the Spider evaluation set. During both the training and the formal study, participants reviewed one example at a time on our developed UI (Section~\ref{sec:interface-design}). At the end of each example, participants were asked to {determine if the AI-generated answer was correct or incorrect} and received immediate feedback automatically presented by the platform. Note that the type of explanations presented to them was decided based on their group. The training session was necessary to educate the non-technical participants about the overall task (e.g., what is a database and what is database querying) and help them get familiar with the UI.
Upon completing all 30 examples in the formal study, participants completed the Jian Scale measure again to evaluate any changes in their trust levels. Finally, to allow for a deeper understanding of the participant behaviors, we requested them to provide feedback on the study through an open-ended form, sharing their experience and any issues encountered during the survey.

Our study received the approval from the university's institutional review board (IRB).

\section{Results and Analysis}\label{sec:results-analysis}
\subsection{Performance of Participants with Different Transparency Levels} \label{sec:analysis-performance}

\begin{table*}[t!]
    \centering
    \resizebox{0.85\textwidth}{!}{
    \begin{tabular}{cccccc}
    \toprule
     & \textbf{Overall (\#=97)} & \textbf{Low (\#=30)} & \textbf{Medium (\#=32)} & \textbf{High (\#=35)}  \\
    \toprule 

    Accuracy in Correct AI Predictions (\#=17) & {70.16\%}& {68.82\%} & {68.56\%} & {\textbf{72.77\%}} \\
     \midrule
    Accuracy in Incorrect AI Predictions (\#=13) & {47.66\%}& {\textbf{49.48\%}} & {48.55\%} & {45.27\%} \\
     \midrule 
    Total Accuracy (\#=30) & {60.41\%}& {60.44\%} & {59.89\%} & {\textbf{60.85\%}} \\
     \bottomrule
    \end{tabular}
}
    \caption{The accuracy of participants correctly identifying correct or incorrect AI predictions, when they interacted with explanations at different transparency levels. Within the parentheses, we show the number of samples of each sub-category of performance or sub-group of participants.
    }
    \label{tab: perf-trans}
\end{table*}

We present the participants' performance, i.e., their accuracy in correctly recognizing correct or incorrect model predictions based on the presented explanations {across different transparency levels}, in Table~\ref{tab: perf-trans}. Specifically, for \textbf{Accuracy in Correct AI Predictions}, we considered the subset of 17 test questions where the semantic parser made a correct prediction, and then reported the percentage of predictions that were correctly recognized as correct predictions by the human participants. For \textbf{Accuracy in Incorrect AI Predictions}, we similarly considered a complementary subset of 13 test questions where the semantic parser made an incorrect prediction, and then reported, among these incorrect model predictions, how much portion was correctly recognized as incorrect predictions by the participants. Finally, \textbf{Total Accuracy} presents the overall accuracy across all model predictions of the 30 test questions, no matter if they were correct or incorrect. These metrics are formally described below.

{\small
\begin{align}
    \text{Correct AI Pred Acc} &= \frac{\text{\# of Correct AI Pred Recognized to be Correct}}{\text{\# of Correct AI Pred}}\\
    \text{Incorrect AI Pred Acc} &= \frac{\text{\# of Incorrect AI Pred Recognized to be Incorrect}}{\text{\# of Incorrect AI Pred}}\\
    \text{Total Acc} &= \frac{\text{\# of Correctly Recognized AI Pred}}{\text{\# of AI Pred}}
\end{align}
}

For each subset of model predictions, Table~\ref{tab: perf-trans} also presents the subgroup performance depending on the type of model explanations that a participant interacted with. We discuss the main findings below.

\paragraph{Participants interacting with explanations at different transparency levels had a similar overall performance.} The total accuracy of participants at identifying whether the model predictions were correct or incorrect was 60.41\%, as shown in Table~\ref{tab: perf-trans}. This total accuracy remained similar across different transparency levels, suggesting that different transparency levels of explanations had little effect on participants' total performance in distinguishing between correct and incorrect model predictions.

\paragraph{Participants were less successful in identifying incorrect AI predictions.} 
{When breaking down the accuracy by correct and incorrect subsets of model predictions, participants were significantly more accurate when the model prediction was correct (70.16\% vs. 47.66\% total accuracy). This indicates a general tendency of participants to accept the AI prediction, a pattern observed across all transparency levels, with the effect being most pronounced among high-transparency participants. The provided explanation 
could be partially responsible for biasing the participants to place greater confidence in the AI's predictions, an observation consistent with prior works~\cite{buccinca2021trust, lai2019human}. Given that high-transparency explanations offer more detailed insights into the AI’s decision-making process, they may have been more persuasive. On the other hand, considering that the participants are all users without technical or AI backgrounds, the huge amount of information included in the high-transparency explanations may be cognitively too much for them, making them unable to fully digest the information or skip the information when making the judgment. In contrast, participants interacting with low-transparency explanations tend to be more cautious, demonstrated by their lower accuracy in correct AI predictions and higher accuracy in incorrect ones, given that the participants were only provided with model confidence as the explanation.}

\begin{figure}[t]
    \centering
    \begin{subfigure}[b]{0.45\textwidth}
        \centering
        \includegraphics[width=\textwidth]{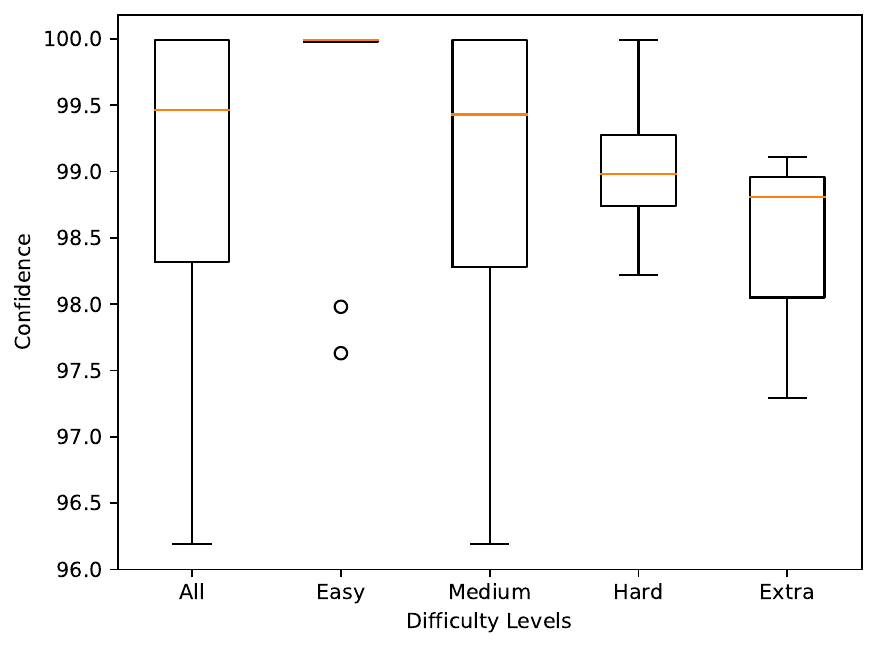} 
        \caption{Model Confidence Distribution}
        \label{fig:confidence}
    \end{subfigure}
    \hfill
    \begin{subfigure}[b]{0.45\textwidth}
        \centering
        \includegraphics[width=\textwidth]{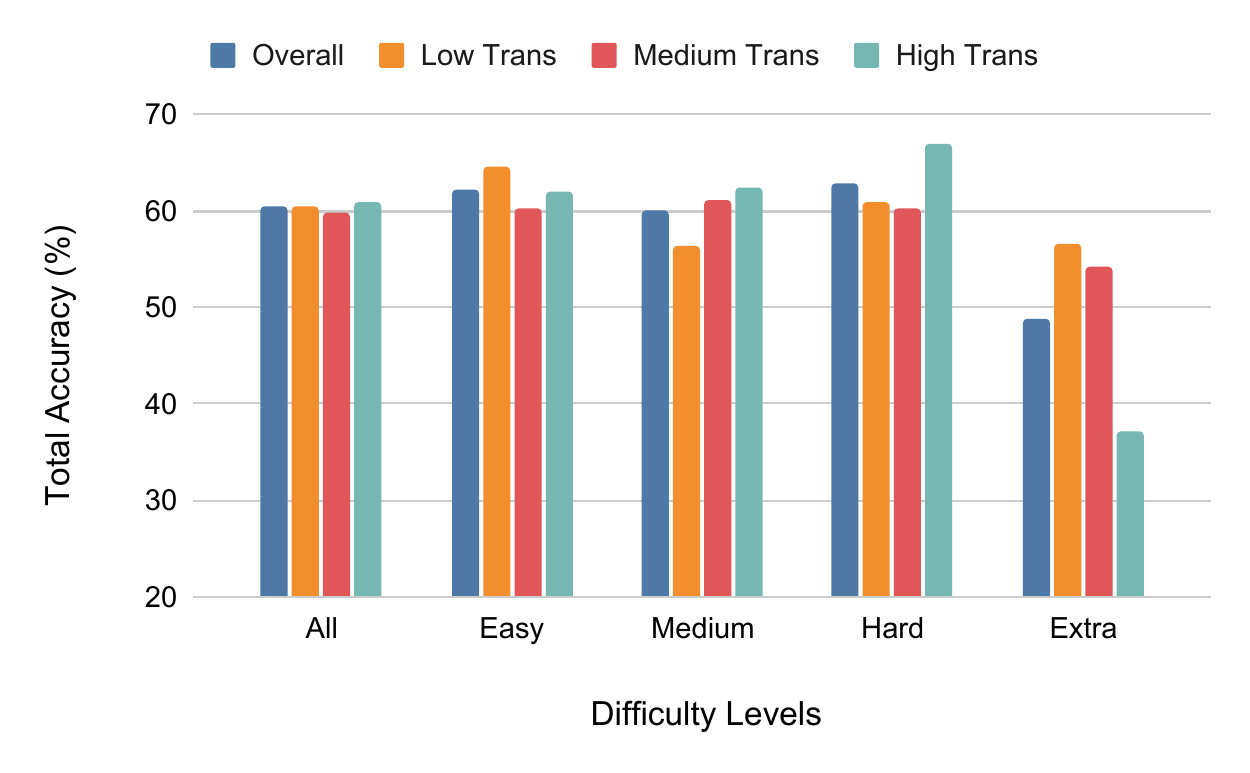} 
        \caption{Participant Group's Total Accuracy across Difficulty Levels}
        \label{fig:trans-diff-perf}
    \end{subfigure}
    \caption{(a) The distribution of model confidence in the low-transparency explanations across varying task difficulty levels. (b) The Total Accuracy of participants interacting with different transparency levels of explanations on different difficult levels of examples.}
    \label{fig:conf-acc-difficulty-levels}
\end{figure}

To further understand the participant performance, we plot the distribution of the model confidence (as reported in the low-transparency explanations) across different difficulty levels of test questions, and the participants' Total Accuracy in recognizing correct vs. incorrect predictions similarly across different difficulty levels, in Figure~\ref{fig:conf-acc-difficulty-levels}. From Figure~\ref{fig:confidence} we observed that the neural semantic parser used in our experiments tended to be overly confident, no matter if its predictions are actually correct or incorrect. This is consistent with discoveries in prior research~\cite{stengel2023calibrated, jiang2021can, desai2020calibration}. However, between difficulty levels where the model makes substantially fewer (e.g., Easy questions) or more (e.g., Extra Hard questions) incorrect predictions (Figure~\ref{fig:semantic-parser-accuracy}), the model did show a lower confidence score, which explains how low-transparency participants could recognize correct/incorrect model predictions with the highest accuracy in both the Easy and Extra Hard categories (Figure~\ref{fig:trans-diff-perf}).

\paragraph{No consistent impact of transparency levels was observed on the participant performance across difficulty levels, and medium-transparency explanations strike a good balance.} Figure~\ref{fig:trans-diff-perf} also shows that no particular transparency level consistently assisted the participants with the best performance across all difficulty levels of test questions. Low-transparency participants had the best performance on Easy and Extra Hard examples, while high-transparency participants had the best performance on Medium and Hard examples but performed substantially worse than all the other types of explanations on Extra Hard examples. As we discussed before, the much worse performance of participants interacting with high-transparency explanations could be due to that the highly informative high-transparency explanations may have biased the participants to trust in the model predictions, or the participants might not be able to digest this large amount of information, leading to mostly incorrect judgment. Among the three transparency levels of explanations, as shown in both Table~\ref{tab: perf-trans} and Figure~\ref{fig:trans-diff-perf}, the medium-transparency explanations strike a good balance between the low- and high-transparency ones, allowing participants to obtain a balanced accuracy across different subsets of AI predictions and across different task difficulty levels.

\begin{figure}[t!]
    \centering
    \includegraphics[width=0.8\linewidth]{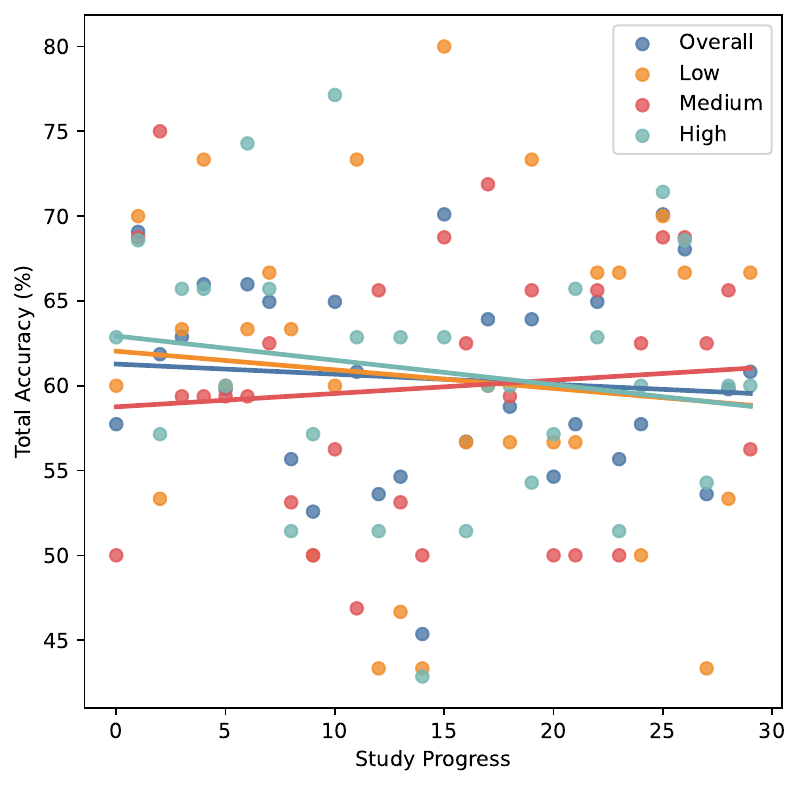}
    \caption{Participants' performance trend over time with different transparency levels of explanations. Regression lines are also plotted.}
    \label{fig:survey_progress_perf}
\end{figure}

\paragraph{Only the medium-transparency explanations allowed for more engaging participant interaction and yielded increasing participant performance over time.}
We further assess whether participants improved their performance over time by gaining a better understanding of the task and the explanations (recall that participants received feedback after their judgment). More importantly, we would like to see if all transparency levels of explanations facilitated this self-adaptation and learning process. To answer this question, in Figure~\ref{fig:survey_progress_perf}, we plot the participants' Total Accuracy calculated for every question ID (from 0 to 29), grouped by the type of explanations they interacted with or showing an overall. For example, the orange dot in the first column of the plot represents an $\sim$0.60 Total Accuracy of low-transparency participants when they worked on their first question, and the highest orange dot in the middle column of the plot shows a Total Accuracy of $\sim$0.80 for the 16th test question that low-transparency participants worked on. We then ran a linear regression model to fit the data points and plotted the regression line.
Note that while the order of test questions each participant received could be different, we expect that, when aggregating their performance, this plot should display a representative line of how they gradually adapt to the model. Surprisingly, we observed that the accuracy of participants overall showed a decreasing trend, suggesting a potential negative effect of explanations on accuracy with continued engagement. We conjecture the reason to be that the participants, who came without technical backgrounds, became increasingly confused and fatigued when they kept interacting with AI in such a complex task. However, viewing accuracy trends across different transparency levels revealed that interestingly medium-transparency participants showed an upward accuracy trend, while those in the low and high-transparency showed a downward trend. This suggests that medium-transparency explanations were more effective, enabling participants to gradually learn to understand the model's decision-making process
and better distinguish between correct and incorrect AI predictions after they became used to the task and explanation.

\subsection{View Time Analysis of Participants}

We tracked participants' view time for each test question during the user study to analyze how much time they needed to understand the prediction and the explanation. Specifically, for each test question, the view time is defined as the time from when the question was displayed to when the participant submitted their response.

\begin{figure}[t!]
    \centering
    \includegraphics[width=0.9\linewidth]{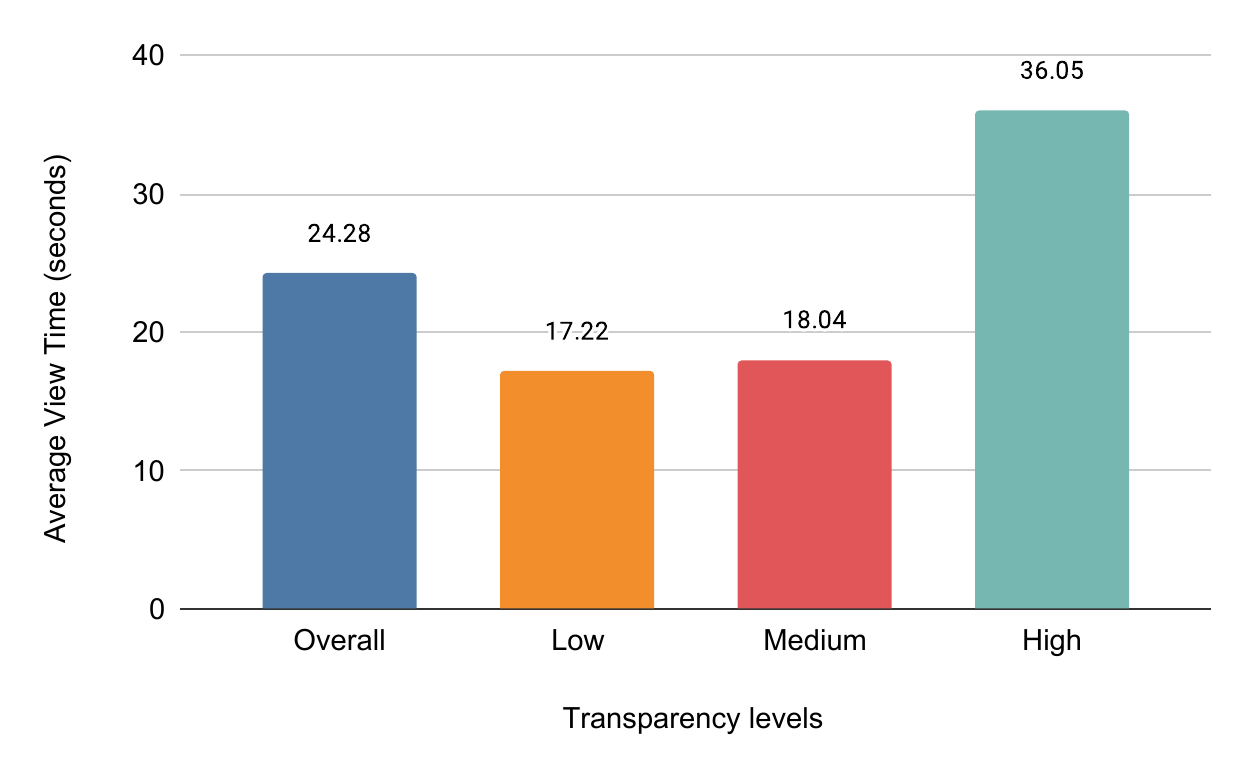}
    \caption{The average view time of participants with different transparency levels of explanations, defined as the time when a test question was displayed till the time when a participant submitted their response.}
    \label{fig:view_time_diff}
\end{figure}

\paragraph{High-transparency explanations are the most time-consuming, and low- and medium-transparency explanations resulted in a similar view time.} The overall average view time per test question was 24.28 seconds, as shown in Figure~\ref{fig:view_time_diff}. As expected, high-transparency participants had the longest average view time of 36.05 seconds, followed by the medium-transparency participants at 18.04 seconds, and the low-transparency participants at 17.22 seconds. Interestingly, the low and medium-transparency participants had a similar view time with a difference of less than a second, suggesting that new information included in the medium-transparency explanation did not significantly increase the cognitive load on participants.

\begin{figure}[t]
    \centering
    \begin{subfigure}[b]{0.38\textwidth}
        \centering
        \includegraphics[width=\textwidth]{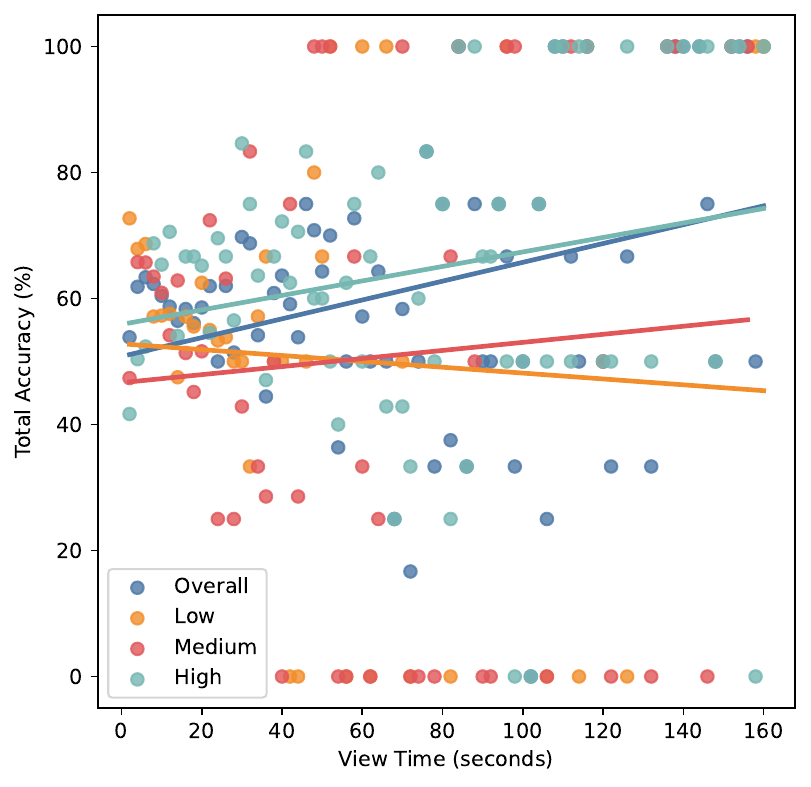}
        \caption{By Transparency Levels of Explanations}
        \label{fig:view_time_perf_trans}
    \end{subfigure}
    \hfill
    \begin{subfigure}[b]{0.38\textwidth}
        \centering
        \includegraphics[width=\textwidth]{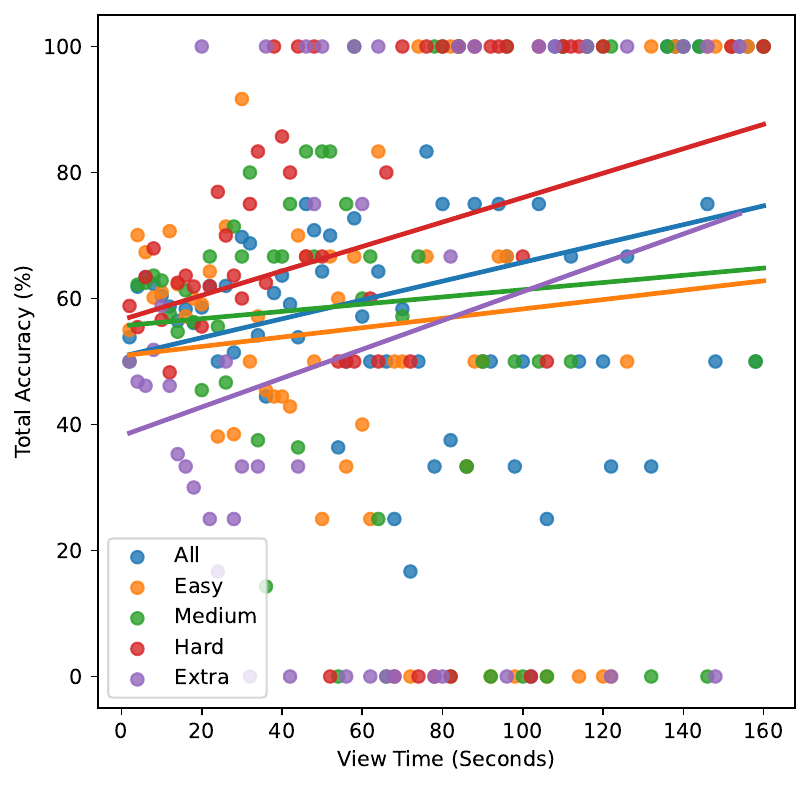} 
        \caption{By Difficulty Levels of Test Questions}
        \label{fig:view_time_diff_perf}
    \end{subfigure}
    \caption{Accuracy trends of participants across (a) different transparency levels of explanations and (b) different difficulty levels of test questions as view time increases.}
    \label{fig:view-time-analysis}
\end{figure}

\paragraph{Longer viewing times led to an increase in accuracy across transparency levels and difficulty levels, except for low-transparency explanations.} We further plot two distributions of participant performance per view time, one by the transparency level of explanations and the other by the difficulty level of questions, in Figure~\ref{fig:view-time-analysis}. Specifically, for the ``Overall'' or ``All'' curve in the two plots, after measuring the view time of each test question from each participant, we grouped the samples into different bins of view time, each in an interval of 2 seconds, and then calculated the Total Accuracy of participant performance for each bin using samples inside the bin. For subgroup results, we based the calculation on the corresponding subset of samples.
We first examine the impact of transparency levels (Figure~\ref{fig:view_time_perf_trans}). As indicated by the ``Overall'' curve, participants who spent more time on each example generally achieved higher accuracy in distinguishing between correct and incorrect model predictions. This was particularly prominent among participants in the medium- and high-transparency groups. This observation indicates that explanations are effective but participants need to take the time to understand them, highlighting the importance of optimizing explanations to enhance user engagement. In contrast, low-transparency participants showed a decreasing trend in accuracy as they spent more viewing time. Since these participants only received the model's confidence score as an explanation, extended viewing time did not provide additional value and may have instead led to confusion.
We also look at the impact of task difficulty levels in Figure~\ref{fig:view_time_diff_perf}. Similar to the first plot, we observed a positive correlation between the participant's Total Accuracy and their per-question view time, and this observation is consistent across all difficulty levels. In addition, the effect was most pronounced for Hard and Extra Hard questions, which indicates that participants needed more time to fully process the explanations for these more challenging questions, and when they did, their accuracy significantly improved.

\begin{figure}[t!]
    \centering
    \includegraphics[width=0.8\linewidth]{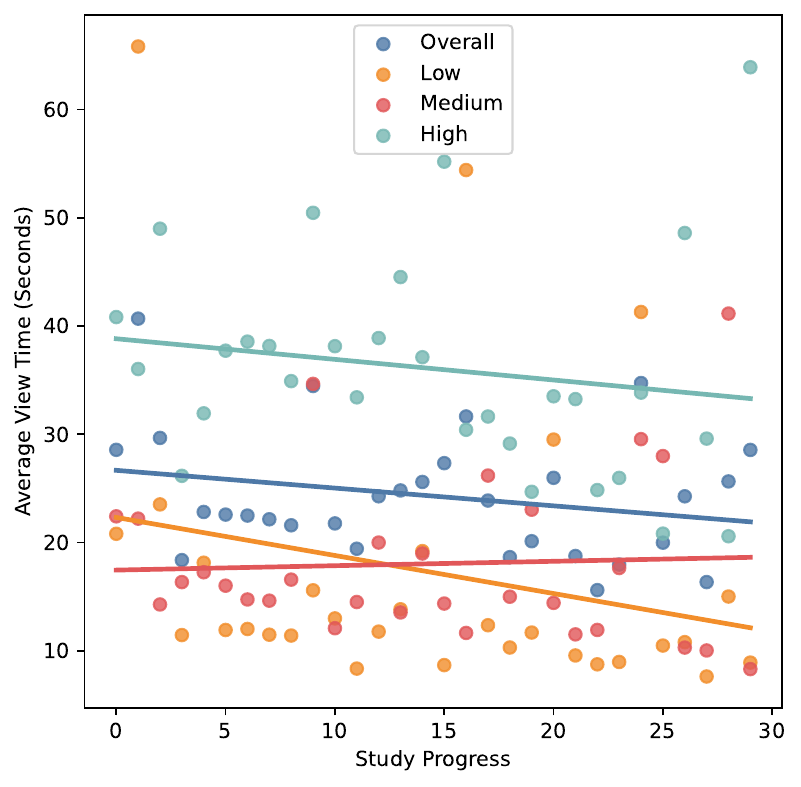}
    \caption{Average view time trends across different transparency levels as the study progressed.}
    \label{fig:survey_progress_view}
\end{figure}

\paragraph{Only the view time for the medium-transparency explanations increased as the study progressed.} 
Following a similar analysis of Figure~\ref{fig:survey_progress_perf}, we show the changes in participants' average view time per test question over time in Figure~\ref{fig:survey_progress_view}. Interestingly, we see that as the study proceeded, participants interacting with low- and high-transparency explanations spent less time on each test question, suggesting that these participants likely stopped engaging with the explanation over time.
In contrast, the increase in view time for medium-transparency participants suggests that they progressively understood and were able to engage better with the explanations. These observations confirmed our conjectures with Figure~\ref{fig:survey_progress_perf} and explained why only medium-transparency participants exhibited increasing capability in correctly recognizing predictions over time.
The finding also underscores that neither minimal nor excessive explanations are optimal; rather, providing a balanced amount of information improves user engagement and leads to more effective use of AI models.

\subsection{Trust Analysis of Participants}
As discussed in Section~\ref{sec:trust-measurement}, we measured participants' dispositional trust using the propensity to trust measure and their learned trust using the Jian Scale.
Below, we analyze these two trust measures across participants receiving varying levels of algorithm transparency.

\begin{figure}[t]
    \centering
    \begin{subfigure}[b]{0.38\textwidth}
        \centering
        \includegraphics[width=\textwidth]{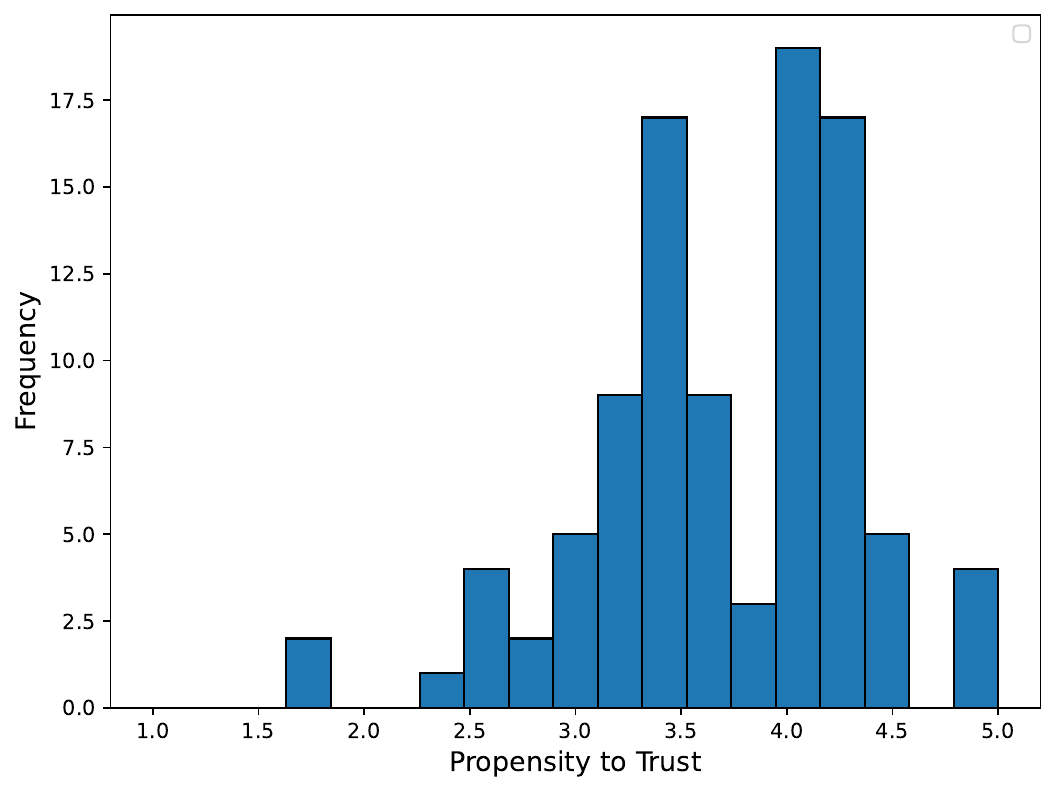}
        \caption{Propensity to Trust Distribution}
        \label{fig:prop_to_trust_dist}
    \end{subfigure}
    \hfill
    \begin{subfigure}[b]{0.38\textwidth}
        \centering
        \includegraphics[width=\textwidth]{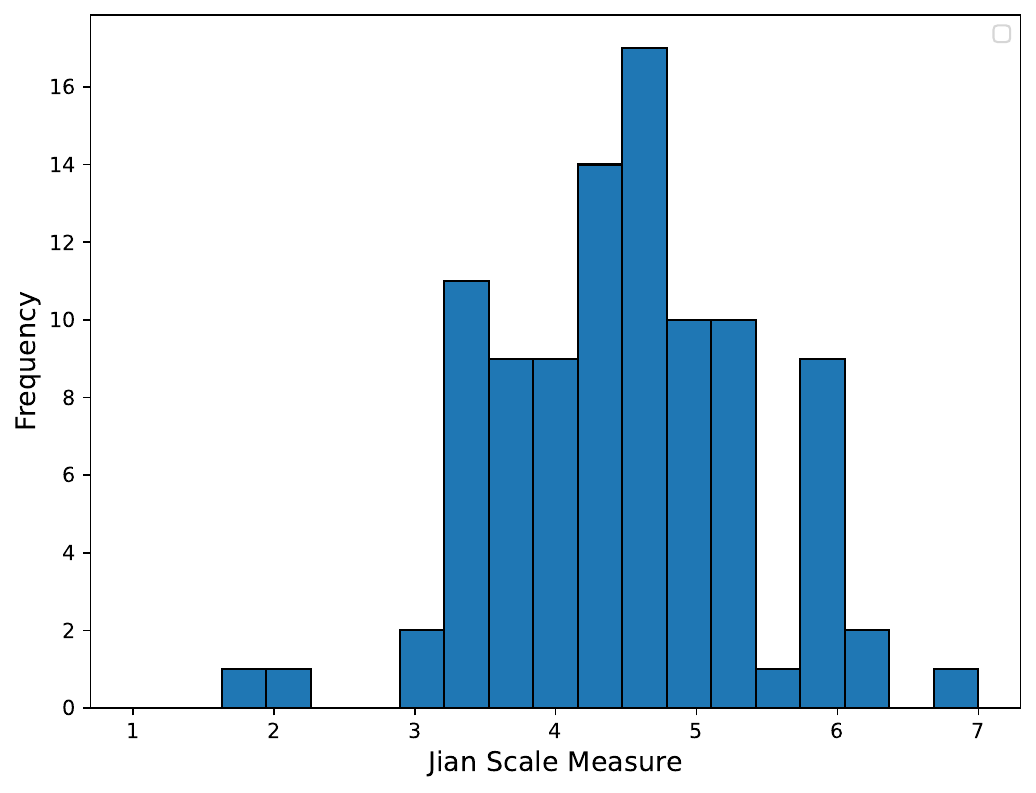} 
        \caption{Jian Scale Measure Distribution}
        \label{fig:jian_scale_measure_dist}
    \end{subfigure}
    \caption{The distribution of the participants (a) Propensity to Trust and (b) Jian Scale Measure scores before the study.}
    \label{fig:trust-measure-dist}
\end{figure}

\paragraph{Jian Scale measure and propensity to trust have a positive correlation.} In Figure~\ref{fig:trust-measure-dist}, we show the distributions of the participants' Jian Scale and propensity to trust scores before the study. For Jian Scale measurement, because the first five questions ask about negative opinions (i.e., the higher the score, the less trust in the model), we flipped the participants' scores from the initial scale of 1-7 to 7-1, so we can report a single average score across all Jian Scale questions. A similar preprocessing was applied to the propensity to trust scores as well. The mean and standard deviation (SD) for the propensity to trust measure were 3.69 and 0.61, respectively, while the mean and SD for the Jian Scale measure were 4.49 and 0.90, respectively. We note that the propensity to trust and Jian Scale measure before the study exhibited a statistically significant positive correlation, with a Pearson coefficient of 0.421 and a p-value of 0.000016. 


\begin{figure}[t!]
    \centering
    \begin{subfigure}[b]{0.45\textwidth}
        \centering
        \includegraphics[width=\textwidth]{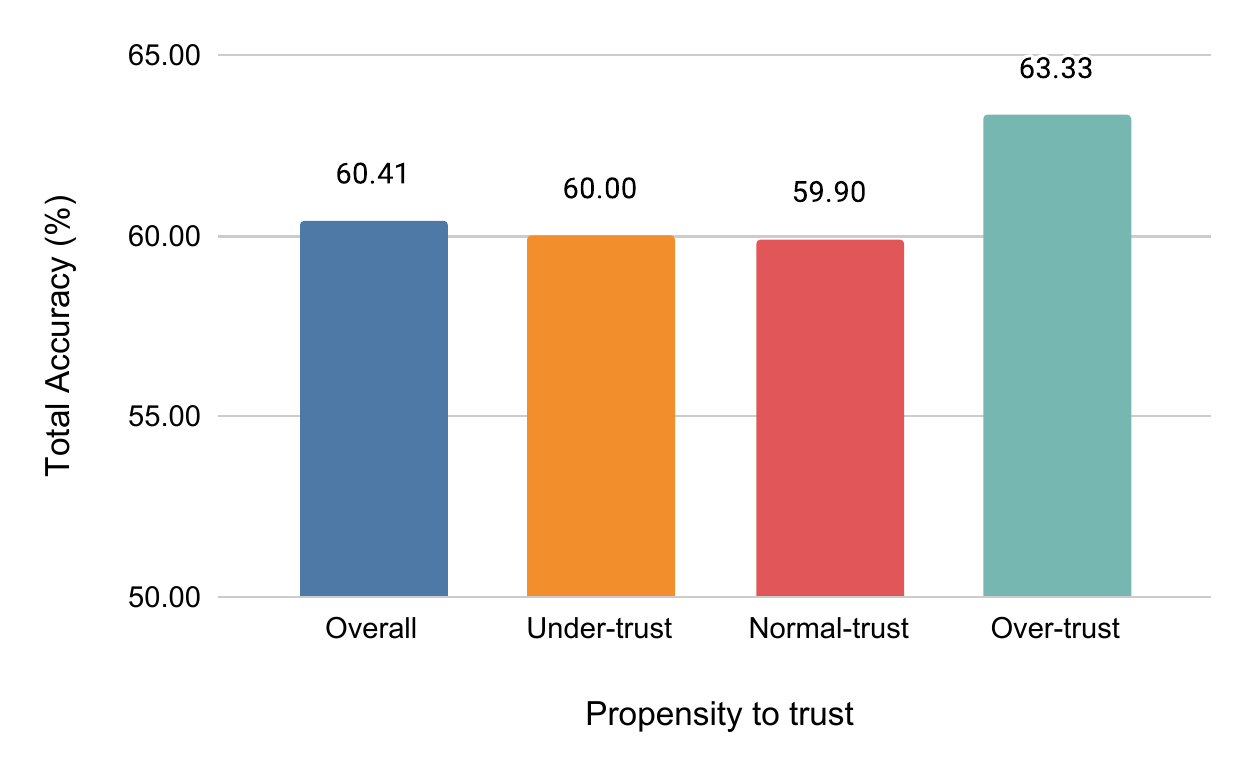} 
        \caption{Participant Total Accuracy per Trust Group}
        \label{fig:prop-to-trust-perf}
    \end{subfigure}
    \hfill
    \begin{subfigure}[b]{0.37\textwidth}
        \centering
        \includegraphics[width=\textwidth]{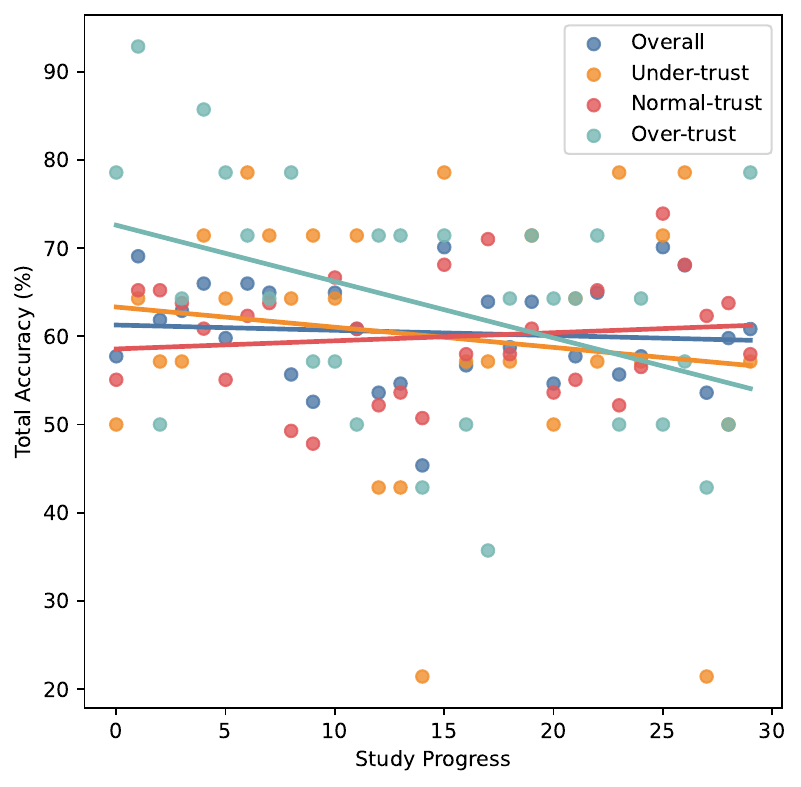} 
        \caption{Participant Performance Change per Trust Group}
        \label{fig:propensity_trend1}
    \end{subfigure}
    \caption{(a) Participant performance grouped by their propensity to trust scores. The three groups of under/normal/over-trust are comparatively defined. No statistically significant difference was observed between groups. (b) Changes in participant performance grouped by their initial trust level.}
    \label{fig:main}
\end{figure}

\paragraph{No significant impact was observed from a participant's dispositional trust to their performance.}
Utilizing the propensity to trust metric, we categorize the participants into three groups -- Under-trust, Normal-trust, and Over-trust, based on their dispositional trust. Specifically,
participants whose trust scores were one standard deviation (0.61) below and above the mean (3.69) were classified as Under-trust and Over-trust, respectively, with the remaining participants categorized as normal-trust. This resulted in 14 participants in the Under-trust group, 14 in the Over-trust group, and 69 in the Normal-trust group. We note that there does not exist any standard that aligns an absolute trust score to a trust group, and our categorization here is a consequence of comparison.
Based on this categorization, we report participants' Total Accuracy per trust group in Figure~\ref{fig:prop-to-trust-perf}. Our t-test result shows no statistically significant difference between groups. The results thus imply that the dispositional trust of a participant does not have a significant impact on their performance.

\paragraph{Only participants starting with a normal dispositional trust showed increasing performance over time.} In Figure~\ref{fig:propensity_trend1}, we further visualize the participant performance of each trust group over time. We observe that only the participants with normal trust showed an increasing trend. This highlights the importance of starting with a normal level of trust as participants become more familiar with the task.

\begin{figure}[t!]
    \centering
    \includegraphics[width=0.9\linewidth]{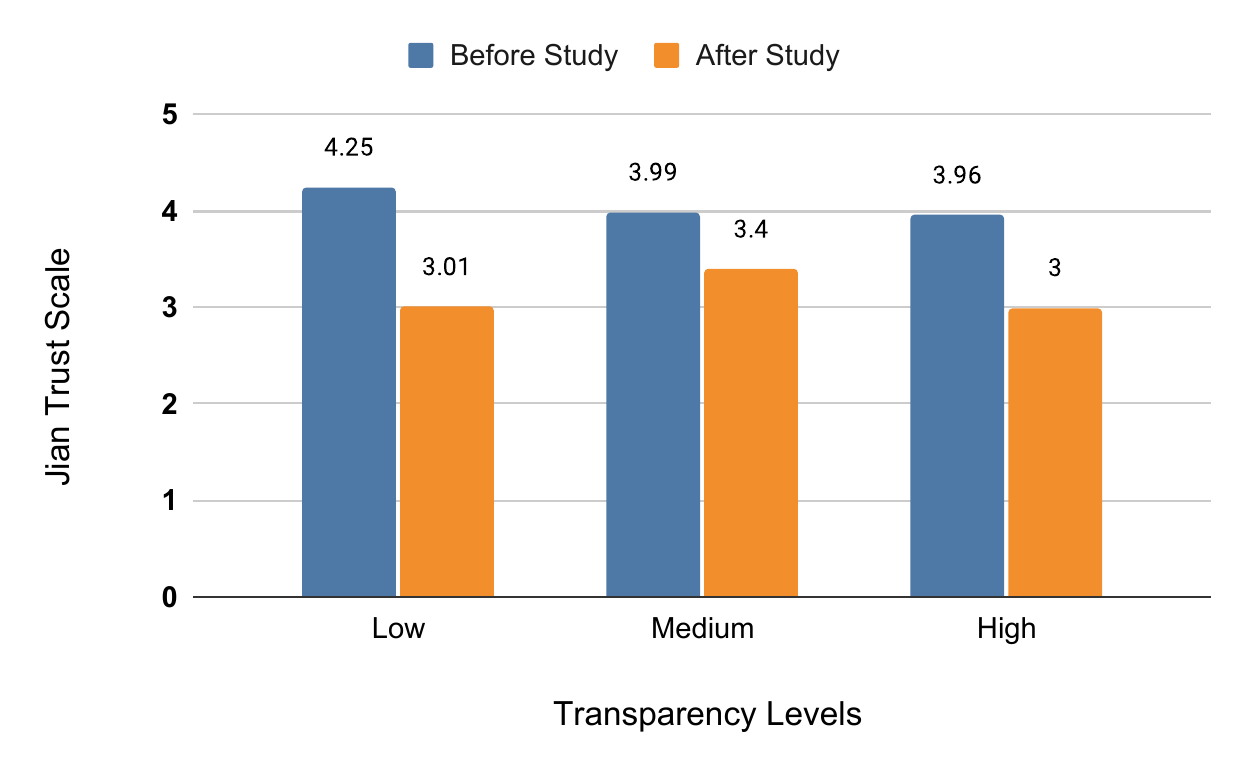}
    \caption{Participants' Jian Scale scores before and after the study, grouped by the transparency levels of their explanations.}
    \label{fig:before-after-jian}
\end{figure}

\paragraph{The trust level of participants decreased after the study, with medium-transparency participants showing the smallest changes.} To assess the change in participants' trust level, the participants were required to fill out the Jian Scale trust measure before and after the study. 
As shown in ~\ref{fig:before-after-jian}, the overall trust level of the participants after the study was decreased. This suggests that participants recognized their tendency to over-trust the AI model and felt the need to reduce their trust. This observation supports the findings in Section~\ref{sec:analysis-performance} on participants' over-trust in AI, where participants demonstrated significantly lower Total accuracy when identifying incorrect AI predictions compared to correct AI predictions.
We also notice that, among the three transparent levels of explanations, the medium-transparency level exhibited the smallest effect on the participants' trust. This indicates that although medium-transparency participants lowered their trust levels, the explanations provided might have been sufficiently reasonable, leading to a more balanced level of trust after the study.

\subsection{Qualitative Analysis of Participants' Post-Study Feedback}
To gain a better understanding of the participants' experience or any issues encountered during the study, they were asked to provide written, fully open-ended feedback at the end of the study. From the collected feedback, we did not identify any situation where participants completed the study without understanding its task setting. We provide the most common feedback below.

\paragraph{AI made mistakes with high confidence scores.} Several participants noted that the model often displayed high confidence, even when it made mistakes, which they found misleading. One participant, for example, remarked, \textit{``It was interesting to see how the system was 99.99\% certain, but was actually incorrect''}. This was expected, as the majority of examples had high confidence levels, with all above 95\%, as shown in Figure~\ref{fig:confidence}. Consequently, the model's confidence scores are an unreliable indicator for predicting its actual accuracy, as noted by one of the low transparency participants, \textit{``It was hard to rely on the system to know if it was correct or not if I didn't know the answer to the question.''} As we discussed in Section~\ref{sec:analysis-performance}, low-transparency participants were thus more conservative in accepting the model predictions.

\paragraph{High-transparency participants found the explanations to be complicated.} High transparency participants reported that the explanations were long and complicated: \textit{``With some of the very long steps to read through, it was complicated to find out which results were correct or incorrect''}. However, some participants also found the explanations to be educational as they provided more insights into the model's decision-making process: \textit{``I think the survey was very educational while still being challenging''}. The ambiguous effect of high-transparency explanations explains their strength in helping participants with Hard test questions and weakness in helping them with Extra Hard questions (Figure~\ref{fig:conf-acc-difficulty-levels}).

\paragraph{The provided explanation was convincing even when the AI made incorrect predictions.} As discussed in Section~\ref{sec:analysis-performance}, explanations could appear to persuade participants that the predictions were correct, even when they were not. One medium-transparency participant remarked, \textit{``The AI was often wrong but still convincing, which is concerning''}.

\section{Discussion: How to Design User-Centered AI Explanations?}\label{sec:discussion}

Explanations for AI predictions should help users determine when and when not to trust the AI's prediction, particularly in high-stakes applications where errors can be costly and issues of safety and bias are critical. If explanations can reliably signal when to trust or challenge AI predictions, AI adoption in various domains could be safer, even if the models are not always accurate.
However, our study discovered that when the task is complicated yet the users lack sufficient backgrounds, an improperly designed explanation (e.g., the high-transparency explanation) could persuade users to accept AI predictions without sufficient scrutiny. This discovery implies a pressing need for more advanced approaches to design AI explanations that facilitate user understanding and avoid misleading. Below, we discuss a list of possible solutions:

\paragraph{Counterfactual questions as explanations.}
We argue that developing explanation methods that help users identify incorrect AI predictions is essential. One possible solution is to automatically generate or guide users to compose counterfactual questions based on their initial questions. By comparing the AI predictions to the counterfactual questions and the prediction to the original question, users may find it easier to judge the correctness of the AI result. For example, to verify an AI's response to the question ``List the names of professors in the CS department above age 56'', a human participant or another AI model could write or generate follow-up or counterfactual questions such as ``List the names of professors in the CS department'' or ``List the names and ages of professors in the CS department''. Users can then cross-check the consistency of the responses from all the questions when deciding if the prediction to the original one is correct.

\paragraph{Interactive and user-specific explanations.}
We observed that providing too much or too little explanation about the AI's decision-making process leads to a decline in overall accuracy and participants' engagement with the explanations over time. This emphasizes the importance of balancing the cognitive load of participants by providing just enough information for participants to evaluate AI predictions effectively. Thus, a more effective strategy may involve an interactive approach, where explanations are provided incrementally based on participants' needs, rather than overwhelming them with extensive details upfront. In addition, the explanations can also be optimized to make them more engaging by including different visual aids and interactive elements.

\paragraph{Dynamic explanation types that calibrate human-AI trust.}
A user's trust in AI is a key factor in determining whether to rely on an AI’s predictions or not. Over-trust can result in users blindly accepting AI outputs, while under-trust can cause excessive skepticism and inefficient use of machine power. An ideal system should foster appropriate trust, where users accept the AI's predictions when they are correct and remain cautious when the predictions are incorrect. To achieve this goal, researchers can look into approaches for automatically detecting real-time human-AI trust, followed by mechanisms to dynamically adjust the explanation methods to dampen over-trust and repair under-trust (called ``trust calibration''~\cite{de2018automation, de2020towards}).

\paragraph{User performance as a reward for learning to explain.}
Finally, our study results suggest deeper collaborations between AI and HCI, such that the impact of an explanation method could be accurately measured against prospective users in an early stage, and this real impact can be converted into a reward that steers the development of the explanation method. However, this solution still faces multiple challenges, including how to design the automatic reward function, whether it is feasible to fit the user performance with such a function, and how the explanation method development can leverage the reward outcome. These challenges could be particularly prominent when the AI task is complicated (e.g., semantic parsing).

\section{Conclusion}
In this study, we conducted a non-expert human-subject experiment on the text-to-SQL semantic parsing task to investigate how explanations at three levels of algorithm transparency impact humans' capability in recognizing correct vs. incorrect AI predictions and human-AI trust. While previous research addressed similar questions with simpler classification tasks, it was unclear if those insights would apply to more complex tasks (e.g., structured prediction, exemplified by semantic parsing) involving non-experts. 
Our study discovered multiple interesting findings, with the most prominent one being the better effectiveness of medium-transparency explanations than explanations at low and high levels. This effectiveness was demonstrated by participants' increasing view time and improving performance, as well as their relatively less reduction in trust in AI, when interacting with this type of explanation.
These findings should be carefully considered by future researchers when designing model explanation methods or employing AI in high-stakes decision-making processes.

\begin{acks}
This project was sponsored by Virginia's Commonwealth Cyber Initiative, National Science Foundation (SHF 2311468/2423813), and the College of Computing and Engineering and the Department of Computer Science at George Mason University. This project was also supported by resources provided by the Office of Research Computing at George Mason University (https://orc.gmu.edu) and funded in part by grants from the National Science Foundation (2018631).
\end{acks}


\bibliographystyle{ACM-Reference-Format}
\bibliography{main}


\end{document}